\documentclass{jpp-aasmacros}
\usepackage{graphicx}
\usepackage{epstopdf, epsfig}
\usepackage{hyperref}
\usepackage{amsmath,amsfonts,amssymb}
\usepackage{subfigure}
\usepackage{booktabs}
\usepackage[normalem]{ulem}  
\usepackage{bm}  

\newcommand*{\mt}{\mathrm}
\newcommand*{\unit}[1]{\;\mt{#1}}  
\newcommand*{\abt}{\mathord{\sim}} 
\newcommand*{\ptl}{\partial}
\newcommand*{\dtl}{\mathrm{d}}

\newcommand*{\del}{\nabla}
\renewcommand{\vec}[1]{\boldsymbol{#1}}  

\newcommand*{\ptlff}[2]{\frac{\partial #1}{\partial #2}}

\newcommand*{\prll}{\parallel}

\newcommand*{\me}{m_{\mathrm{e}}}  
\newcommand*{\mi}{m_{\mathrm{i}}}

\newcommand*{\Te}{T_{\mathrm{e}}}  
\newcommand*{\Ti}{T_{\mathrm{i}}}
\newcommand*{\Tio}{T_{\mathrm{i0}}}  

\newcommand*{\Omce}{\Omega_{\mathrm{e}}}  
\newcommand*{\Omci}{\Omega_{\mathrm{i}}}
\newcommand*{\Omcs}{\Omega_{\mathrm{s}}}
\newcommand*{\ompe}{\omega_{\mathrm{pe}}}
\newcommand*{\ompi}{\omega_{\mathrm{pi}}}
\newcommand*{\omps}{\omega_{\mathrm{p}s}}  

\newcommand*{\rLi}{\rho_{\mathrm{i}}}
\newcommand*{\rLs}{\rho_s}  

\newcommand*{\Omcio}{\Omega_{\mathrm{i0}}}
\newcommand*{\rLio}{\rho_{\mathrm{i0}}}
\newcommand*{\lde}{\lambda_\mathrm{De}}

\newcommand*{\vtio}{v_\mathrm{ti0}}  

\newcommand*{\mratio}{R_\mt{m}}  
\newcommand*{\tbounce}{\tau_\mt{bounce}}  
\newcommand*{\tloss}{\tau_\mt{p}}  

\shorttitle{DCLC in WHAM (draft, \today)}
\shortauthor{A.~Tran, S.~J.~Frank, A.~Y.~Le, et al. (draft, \today)}

\title{Drift-cyclotron loss-cone instability in 3D simulations of a sloshing-ion simple mirror}

\author{
Aaron Tran\aff{1}\corresp{\email{atran@physics.wisc.edu}},  
Samuel J.~Frank\aff{2},  
Ari Y.~Le\aff{3},  
Adam J.~Stanier\aff{3},  
Blake A.~Wetherton\aff{3},  
Jan Egedal\aff{1},  
Douglass A.~Endrizzi\aff{2},  
Robert W.~Harvey\aff{4},
Yuri V.~Petrov\aff{4},  
Tony M.~Qian\aff{1,5},  
Kunal Sanwalka\aff{1},  
Jesse Viola\aff{2},  
Cary B.~Forest\aff{1,2}, 
and Ellen G.~Zweibel\aff{1,6}  
}

\affiliation{
\aff{1}Department of Physics, University of Wisconsin--Madison, Madison, WI, USA
\aff{2}Realta Fusion, Madison, WI, USA
\aff{3}Los Alamos National Laboratory, Los Alamos, NM, USA
\aff{4}CompX, Del Mar, CA, USA
\aff{5}Princeton Plasma Physics Laboratory, Princeton, NJ, USA
\aff{6}Department of Astronomy, University of Wisconsin--Madison, Madison, WI, USA
}

\begin{document}

\maketitle

\begin{abstract}
The kinetic stability of collisionless, sloshing beam-ion ($45^\circ$ pitch
angle) plasma is studied in a 3D simple magnetic mirror, mimicking the
Wisconsin High-temperature superconductor Axisymmetric Mirror (WHAM)
experiment.
The collisional Fokker-Planck code CQL3D-m provides a slowing-down beam-ion
distribution to initialize the kinetic-ion/fluid-electron code Hybrid-VPIC,
which then simulates free plasma decay without external heating or fueling.
Over $1$--$10\unit{\mu s}$, drift-cyclotron loss-cone (DCLC) modes grow and
saturate in amplitude.
DCLC scatters ions to a marginally-stable distribution with gas-dynamic rather
than classical-mirror confinement.
Sloshing ions can trap cool (low-energy) ions in an electrostatic potential
well to stabilize DCLC, but DCLC itself does not scatter sloshing beam-ions
into said well.
Instead, cool ions must come from external sources such as charge-exchange
collisions with a low-density neutral population.
Manually adding cool $\abt 1\unit{keV}$ ions improves beam-ion confinement
several-fold in Hybrid-VPIC simulations, which qualitatively corroborates prior
measurements from real mirror devices with sloshing ions.
\end{abstract}


\section{Introduction}

The Wisconsin High-temperature superconductor Axisymmetric Mirror (WHAM) is a
new laboratory experiment that confines hot plasmas in a magnetic mirror with a
maximum field of $17\unit{T}$ on axis, generated by high-temperature
superconductors (HTS) \citep{endrizzi2023}.
For WHAM and future mirror devices \citep{simonen2008,bagryansky2024,forest2024} to succeed,
both fluid and kinetic plasma instabilities must be quelled.

A kinetic instability of particular concern is the drift-cyclotron loss-cone
(DCLC) instability \citep{post1966,baldwin1977}.
DCLC comprises a spectrum of ion Bernstein waves, coupled to a collisionless
drift wave, that is excited by a spatial density gradient $\del n$ and a
loss-cone ion velocity distribution.
In a magnetized plasma column, DCLC appears as an electrostatic wave that
propagates around the column's azimuth in the ion diamagnetic drift direction,
perpendicular to both $\vec{B}$ and $\del n$.
DCLC can be unstable solely due to $\del n$ when the gradient length scale
$n/(\del n)$ is of order the ion Larmor radius $\rho_\mt{i}$, even for
distributions without a loss cone (e.g., Maxwellians), in which case it may be
called drift-cyclotron instability \citep{mikhailovskii1963}.
In this manuscript, we call both drift-cyclotron and drift-cyclotron loss-cone
modes by ``DCLC'' for simplicity.

Many mirror devices have measured electric and/or magnetic fluctuations at
discrete ion cyclotron harmonics having properties consistent with DCLC.
These devices include
PR-6 \citep{bajborodov1971,ioffe1975},
PR-8 \citep{piterskii1995},
2XIIB \citep{coensgen1975},
TMX and TMX-U \citep{drake1981,simonen1983,berzins1987},
LAMEX \citep{ferron1984},
MIX-1 \citep{koepke1986-BRLD,koepke1986-3D,mccarrick1987,burkhart1989},
GAMMA-6A \citep{yamaguchi1996},
and GDT \citep{prikhodko2018,shmigelsky2024-dclc}.
Experiments on these devices showed that DCLC may be partly or wholly
stabilized by filling the ions' velocity-space loss cone
via axial plasma stream injection
\citep{ioffe1975,coensgen1975,kanaev1979,correll1980,drake1981,simonen1983,berzins1987},
filling the loss cone via angled neutral beam injection, which creates
a non-monotonic axial potential that traps cool ions
\citep{kesner1973,kesner1980,fowler2017,shmigelsky2024-dclc},
decreasing $\del n$ with respect to the ion Larmor radius $\rho_\mt{i}$
\citep{ferron1984},
and bounce-resonant electron Landau damping \citep{koepke1986-BRLD}.
Other effects theoretically calculated to modify and/or help stabilize DCLC
include finite plasma beta \citep{tang1972}, radial ambipolar electric fields
\citep{chaudhry1983,sanuki1986}, and low-frequency external electric fields
\citep{aamodt1977-electron,hasegawa1978}.

WHAM's plasma column is a few to several ion Larmor radii ($\rho_\mt{i}$)
in width and so may excite DCLC.
How will DCLC appear in WHAM; i.e., what will be its azimuthal mode number,
oscillation frequency, and amplitude?
Sloshing ions, injected at $45^\circ$ pitch-angle, helped to suppress DCLC in TMX-U endplugs and
are also used on WHAM;
to what extent can sloshing ions similarly suppress DCLC in WHAM?
In general,
how should WHAM's plasma properties be tuned to suppress DCLC?
These questions have been addressed to varying degrees, for previous devices,
via linear theory
\citep{post1966,tang1972,gerver1976,lindgren1976,baldwin1977,cohen1982,cohen1983,ferraro1987,kotelnikov2017,kotelnikov2018},
quasi-linear theory
\citep{baldwin1976,berk1977},
non-linear theory
\citep{aamodt1977-electron,aamodt1977-shift,myer1980},
and 1D and 2D kinetic computer simulations
\citep{cohen1980,aamodt1981,cohen1982,cohen1983,cohen1984,rose2006}.

Here, we address the aforementioned questions using 3D full-device computer
simulations of DCLC growth and saturation in a hybrid (kinetic ion, fluid
electron) plasma model.
Our simulation accounts for many physical effects relevant to WHAM---magnetic geometry,
beam-ion distributions, both radial and axial electrostatic potentials, and
diamagnetic field response---to obtain a fuller and more integrated kinetic
model than was possible decades ago.
We highlight that the initial beam-ion distributions are obtained from a
Fokker-Planck collisional-transport model of a WHAM shot's full $20 \unit{ms}$
duration.
The Fokker-Planck modeling and code-coupling methods are presented by a
companion study, \citet{frank2024}, within this journal issue.

In {\S}\ref{sec:methods} we describe our simulation methods and parameters.
In {\S}\ref{sec:results-overview} to {\S}\ref{sec:scattering},
we characterize three fiducial simulations evolved to $6 \unit{\mu s}$ that
have reached a steady-state decay.
The main instability in all simulations is described and identified as DCLC,
with the aid of an approximate linear dispersion relation for electrostatic
waves in an inhomogeneous, low-$\beta$ planar-slab plasma.
In {\S}\ref{sec:confinement}, particle confinement is shown to obey a ``gas
dynamic'' rather than ``collisionless mirror'' scaling with mirror ratio and
device length.
In {\S}\ref{sec:cool}--\ref{sec:ELD},
we survey well-known ways to stabilize DCLC that may be relevant to WHAM and next-step mirror devices.
We particularly focus on stabilization via trapped cool ions, and we show that
adding cool ions can improve beam-ion confinement by a factor of $2$--$5\times$
in our simulations.
In {\S}\ref{sec:other-modes}--\ref{sec:comparison}
we briefly discuss how DCLC in WHAM fits into a broader landscape of other instabilities and devices.
Finally, {\S}\ref{sec:conclusions} concludes.

\section{Methods} \label{sec:methods}

\begin{figure*}
    \centering
    \includegraphics[width=\textwidth]{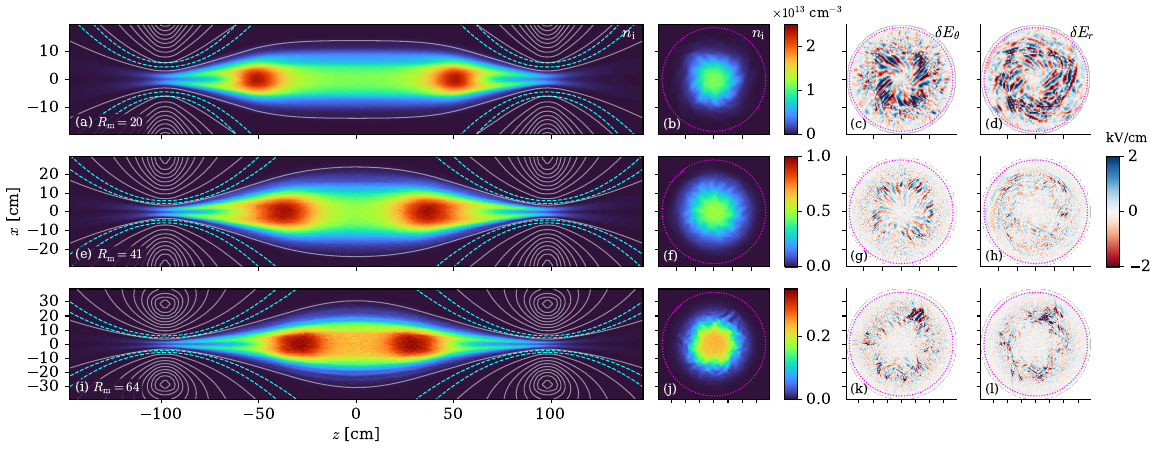}
    \caption{
        2D images of ion density and electric field fluctuations at
        $t \approx 6 \tbounce \approx 6 \mu s$, for three simulations with
        varying vacuum mirror ratio $\mratio = 20$ (top row), $41$ (middle
        row), $64$ (bottom row).
        (a): ion density $n_\mt{i}$ in units of $10^{13} \unit{cm^{-3}}$, 2D
        slice at $y=0$ in $(x,y,z)$ coordinates.
        White lines trace vacuum magnetic fields; dashed cyan lines trace
        hyperresistive dampers and conducting $E=0$ regions (see text).
        (b): like (a), but 2D slice at the mirror's mid-plane $z=0$ showing
        coherent flute-like fluctuations at the plasma edge.
        (c): azimuthal electric field fluctuation $\delta E_\theta$ in kV/cm;
        magenta dotted line traces radial conducting boundary.
        (d): like (c), but radial fluctuation $\delta E_r$.
        Middle row (e)-(h) and bottom row (i)-(l) are organized like panels
        (a)-(d).
        Aspect ratio is distorted in left column (a),(e),(i);
        aspect ratio is to scale in all other panels.
        The ion bounce time $\tau_\mt{bounce}$
        is defined later in \S\ref{sec:methods-plasma}.
    }
    \label{fig:overview2d}
\end{figure*}

\subsection{Simulation Overview} \label{sec:methods-overview}

We simulate freely-decaying plasma in a 3D magnetic mirror made of one central
cell and two expanders (Figure~\ref{fig:overview2d}, left column).
Three magnetic-field configurations are used, labeled by vacuum mirror ratio
$\mratio = \{20, 41, 64\}$, to span WHAM's operating range.
WHAM's magnetic field is created by two HTS coils at $z = \pm 98 \unit{cm}$
and two copper coils at $z = \pm 20 \unit{cm}$ \citep{endrizzi2023}.
When both HTS and copper coils are fully powered,
the magnetic field on axis varies between
$B \approx 17.3 \unit{T}$ at the mirror throats
to $0.86 \unit{T}$ at the device's center ($\mratio=20$).
When the copper coils are partly powered,
$B$ on axis ranges between $17.2$ to $0.414 \unit{T}$ ($\mratio=41$).
When the copper coils are unpowered,
$B$ on axis ranges between $17.1$ to $0.267 \unit{T}$ ($\mratio=64$).

Our simulations are performed with the code Hybrid-VPIC\footnote{
    Publicly available at
    \url{https://github.com/lanl/vpic-kokkos/tree/hybridVPIC}.
}
\citep{le2023-hyb,bowers2008},
which models ion kinetics using the particle-in-cell (PIC) method
and models electrons as a neutralizing fluid.
Ions are advanced using a Boris pusher
\citep{bowers2008}.
Electric and magnetic fields $\vec{E}$, $\vec{B}$ are evolved on a rectilinear
Cartesian ($x,y,z$) mesh.
Particle-mesh interpolation uses a quadratic-sum shape
\citep[Appendix~B]{le2023-hyb};
no filtering of the deposited particle charge and currents is applied.
The magnetic field is advanced using Faraday's Law,
$\ptl \vec{B}/\ptl t = - c\del \times \vec{E}$, with a 4th-order Runge-Kutta
scheme.
The electric field is passively set by a generalized Ohm's law without electron
inertia:
\begin{equation} \label{eq:ohm}
    \vec{E} = - \frac{\vec{V}_\mt{i} \times \vec{B}}{c}
    + \frac{\vec{j} \times \vec{B}}{e n_\mt{e} c}
    - \frac{\del P_\mt{e}}{e n_\mt{e}}
    + \eta \vec{j}
    - \eta_\mt{H} \del^2 \vec{j}
    \, ,
\end{equation}
assuming both $\vec{j} = c \del \times \vec{B} / (4\pi)$
and $n_\mt{e} = n_\mt{i}$.
We further take $P_e = n_e T_e$, with
$T_e$ constant in time and space (isothermal).
Here $n_\mt{i}$ and $n_\mt{e}$ are ion and electron number densities,
$\vec{V}_\mt{i}$ is bulk ion velocity, $P_\mt{e}$ is scalar electron
presure, $\vec{j}$ is current density, $\eta$ is resistivity, $\eta_\mt{H}$ is
hyper-resistivity, $c$ is the speed of light, and $e$ is the elementary charge.
Gaussian CGS units are used in this manuscript unless otherwise stated.

Coulomb collisions are neglected because the ion-ion deflection and ion-electron
drag timescales in WHAM are of order $\mathcal{O}(\mathrm{ms})$, longer than
our simulation durations $\abt 1$--$10 \unit{\mu s}$.

The hybrid-PIC equations solved here are non-relativistic: the displacement current
$\ptl \vec{E}/\ptl t/(4\pi)$ is omitted from Amp\'{e}re's law, and no Lorentz
factors are used in the Boris push.
The speed of light is effectively infinite.
All code equations are solved in a dimensionless form; the normalizations for
converting code variables into physical units are set by choosing reference
values of density, magnetic field, and ion species' mass and charge.

A density floor of $n_\mt{e} \ge \{15, 6, 1.5\} \times 10^{11} \unit{cm^{-3}}$,
for the $\mratio = \{20, 41, 64\}$ simulations respectively, is applied in
the Hall and ambipolar (pressure gradient)
terms of Equation~\eqref{eq:ohm} to prevent division-by-zero in vacuum and
low-density regions surrounding the plasma.
The density floor is set low enough to obtain the electrostatic potential drop
from $z=0$ out to the mirror throats at $z = \pm 98\unit{cm}$, but
the remaining potential drop from throat into expanders is not captured.
Lowering the density floor increases compute cost, so we sacrifice physics in
the expanders that is less-accurately described by the hybrid-PIC model
anyways.

The simulation timestep $\Delta t$ must be smaller than
a Courant–Friedrichs–Lewy (CFL) limit to resolve grid-scale whistler waves:
$\Delta t \propto n (\Delta z)^2 / B$
for cell size $\Delta z$ less than the ion skin depth.
The density floor in high-$\vec{B}$ vacuum regions thus sets the overall simulation timestep.
The CFL-limited timestep is well below the ion-cyclotron period
and other physical timescales of interest, so we sub-cycle the magnetic-field
update $N_\mt{sub}$ times within each particle push to reduce compute cost.
The CFL limit then applies to $\Delta t/ N_\mt{sub}$,
and larger $\Delta t$ can be used.

We set the resistivity $\eta = 0$ and the hyper-resistivity
$\eta_\mt{H} = 2.75 \times 10^{-14} \unit{s\,cm^2}$.
Hyper-resistivity is used solely to damp high-frequency whistler noise at the
grid scale $k \sim \pi/\Delta z$; $\eta_\mt{H}$ does not represent any sub-grid
physics of interest to us.
The hyper-resistive $\vec{E}$ is included in the ion push, since it is not used
to model electron-ion friction.\footnote{
    If hyper-resistivity were used to model electron-ion friction, and no
    explicit collision operator for ions is used, only the frictionless
    $\vec{E}$ should be used in the ion push
    \citep[][Appendix~A]{stanier2019-pic}.
}
Hyper-resistivity is the only explicit form of numerical dissipation in our simulations.

\subsection{Simulation Geometry} \label{sec:methods-geometry}

The simulation domain for the $\mratio=20$ case is a rectangular box with
extent
$L_x = L_y = 39.2 \unit{cm}$ and
$L_z = 294 \unit{cm}$.
The box is decomposed into a $96^2 \times 384$ Cartesian $(x,y,z)$ mesh with
cell dimensions
$\Delta x = \Delta y = 0.41 \unit{cm}$
and $\Delta z = 0.77 \unit{cm}$.
For analysis and discussion, we project data into usual cylindrical coordinates
$(r,\theta,z)$.
For the $\mratio = 41$ and $64$ cases, the domain is enlarged to
$L_x = L_y = 58.8$ and $78.4 \unit{cm}$ while preserving the mesh cell shape,
so the number of mesh points is $144^2 \times 384$ and $192^2 \times 384$
respectively.
The domain extent truncates the expanders at $z = \pm 147 \unit{cm}$, unlike
the real experiment, wherein a set of staggered biasable rings collects
escaping plasma at $z \sim 190$--$210 \unit{cm}$
\citep{endrizzi2023,qian2023}.

The overall simulation timestep $\Delta t = 7.3 \times 10^{-11} \unit{s}$.
The magnetic-field advance is sub-cycled
$N_\mt{sub} = \{100, 250, 1000\}$
times within
$\Delta t$, for $\mratio = \{20, 41, 64\}$ respectively.

Hyper-resistivity $\eta_\mt{H}$ acts like smoothing and removes grid-scale numerical noise
on the whistler-wave dispersion branch, which would otherwise be undamped in the absence of resistivity or hyper-resistivity.
The value of $\eta_\mt{H}$ must be kept small enough to not artificially smooth real
physical phenomena.
The hyper-resistive diffusion timescale estimated as
$\left[\eta_\mt{H} c^2 / (4\pi L^4)\right]^{-1}$ for an arbitrary lengthscale
$L$ is
$1.4 \times 10^{-8} \unit{s}$ for the transverse grid scale
$L \sim \Delta x$; it is
$600 \unit{\mu s}$ for the ion skin depth
$L \sim c/\omega_{\mathrm{pi}} \sim 6 \unit{cm}$
with $n \sim 3 \times10^{13} \unit{cm^{-3}}$.
We cannot make $\eta_\mt{H}$ much larger because the scale separation between
grid noise and physical phenomena is small;
high-$m$ kinetic modes lie below the ion skin depth.
In Appendix~\ref{app:hypereta}, we present density fluctuation properties
from a three-point scan of $\eta_\mt{H}$; some details (e.g., spectral
bandwidth) are altered, but the main conclusions regarding DCLC are not too
sensitive to our chosen value of $\eta_\mt{H}$.

Particle and field boundary conditions are imposed as follows.
A conducting radial sidewall is placed at
$r = 0.47 L_x$, which is in physical units $\{18.4, 27.6, 36.8\} \unit{cm}$
for $\mratio=\{20,41,64\}$ respectively.
A conducting axial sidewall is placed at
$z = 0.485 L_z = \pm 143 \unit{cm}$.
The HTS coils are also surrounded by both conducting and hyper-resistive
wrapper layers (Figure~\ref{fig:overview2d}, left column, dashed cyan curves).
Within the wrapper layer (between nested dashed cyan curves), the grid-local
value of $\eta_\mt{H}$ used in Ohm's Law (Equation~\eqref{eq:ohm}) is increased
$30\times$ to help suppress numerical noise in high-field, low-density regions.
The ``conducting'' boundary is enforced by setting $\vec{E} = 0$ on the mesh,
which disables $\vec{B}$ field evolution.
Bound charge and image currents within conducting surfaces are not
explicitly modeled.
Particles crossing the Cartesian domain boundaries
($x = \pm L_x/2$, $y = \pm L_y/2$, $z=\pm L_z/2$) are removed from the
simulation.
Boundary conditions are applied to $\vec{E}$ at cell centers in a
nearest-grid-point manner, which may contribute to mesh imprinting;
boundaries might be improved with a cut-cell algorithm or simply higher
grid resolution in future work.

\begin{figure}
    \centering
    \includegraphics[width=\textwidth]{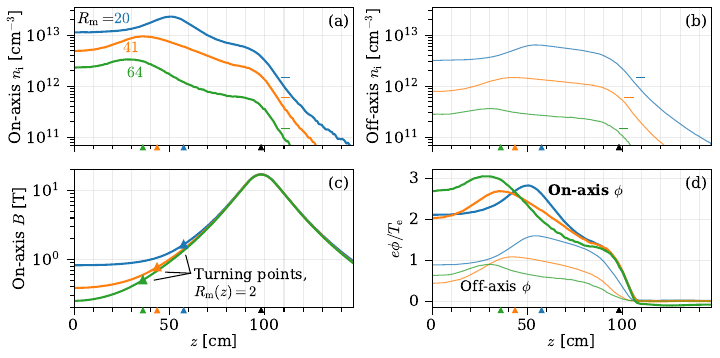}
    \caption{
        Axial profiles of $n_\mt{i}$, $B$, $\phi$ measured at
        $t = 6 \tbounce \approx 6 \mu s$.
        (a): Ion density $n_\mt{i}$ on axis ($r=0$). Dashes mark density floor
        for Ohm's law, Equation~\eqref{eq:ohm}.
        (b): Like (a), but measured along off-axis flux surfaces.
        (c): Magnetic-field strength $B$ on axis.  Ions with $45^\circ$ pitch
        angle turn where the local mirror ratio $\mratio(z)=2$ (triangles).
        (d): Electrostatic potential $e\phi$ in units of electron temperature
        $T_\mt{e}$, measured on-axis (thick curves) and off-axis (thin curves).
        Potentials truncate at $z \sim 100\unit{cm}$, corresponding to density
        floors marked in (a)-(b).
        In all panels: blue, orange, green curves are simulations with vacuum
        $\mratio = \{20, 41, 64\}$ respectively; small triangles mark on-axis
        turning points $\mratio(z) = 2$ (colored) and mirror throat (black).
    }
    \label{fig:overview1d}
\end{figure}

\subsection{Plasma Parameters} \label{sec:methods-plasma}

We model a fully-ionized deuteron-electron plasma
($m_\mt{i} = 3.34\times10^{-24}\unit{g}$)
with typical ion density $n_\mt{i} \sim 10^{12}$ to $10^{13} \unit{cm^{-3}}$
and temperature $T_\mt{i} \sim 5$ to $13 \unit{keV}$
in the mirror's central cell.
The ion velocity distribution is a beam slowing-down distribution with pitch
angle $\cos^{-1}(v_\parallel/v) \sim 45^\circ$ at the mirror mid-plane ($z=0$)
to mimic WHAM's
angled neutral beam injection (NBI).
The beam path is centered on axis ($r=0$).

The ions' spatial and velocity distributions are obtained from the
bounce-averaged, zero-orbit-width, collisional Fokker-Planck code CQL3D-m \citep{petrov2016,forest2024}.
We initialize the CQL3D-m simulations with a $1.5 \times 10^{13} \unit{cm^{-3}}$ plasma at low temperature $T_i = T_e = 250 \unit{eV}$,
mimicking the initial electron-cyclotron heating (ECH) breakdown of a gas puff in WHAM.\footnote{
    The $250\unit{eV}$ temperature is higher than in experiments
    so that we can use coarser velocity-space grid resolution.
    The final evolved solution varies little with initial temperature.
}
The plasma is simulated by CQL3D-m on 32 flux surfaces spanning normalized square root poloidal flux, $\sqrt{\psi_n} = 0.01$--$0.9$, as it is fueled and heated with a realistic $25 \unit{keV}$ neutral beam operating at the experimental parameters.
No heating or fueling sources other than the neutral beam are included.
The velocity-space grid has 300 points in total momentum-per-rest-mass $p/(m c)$,
and either 256 or 300 points in pitch angle.
The total-momentum grid is not linearly spaced, but instead geometrically scaled at low energies to cover the ion distribution function.
The pitch-angle grid is uniformly spaced.
CQL3D-m uses a timestep of $0.0625 \unit{ms}$,
advancing ions and electrons simultaneously.
The neutral beam deposition profile is updated after each timestep using
CQL3D-m's internal FREYA neutral-beam Monte-Carlo solver.
To include the diamagnetic $\vec{B}$-field response to the plasma pressure,
the CQL3D-m solve is iterated with
the MHD equilibrium solver Pleiades\footnote{
    \url{https://github.com/eepeterson/pleiades}
}
\citep{peterson2019-phd},
with improvements to treat pressure-anisotropic equilibria \citep{frank2024}.
CQL3D-m and Pleiades are
coupled using a customized version of the Integrated Plasma Simulator framework \citep{elwasif2010}.
The diamagnetic field is updated in CQL3D-m every $1 \unit{ms}$.

We perform separate CQL3D-m runs for each of the $\mratio=\{20,41,64\}$ cases.
In each case, the NBI power is adjusted in $100 \unit{kW}$ increments, until the $1 \unit{MW}$ maximum input power of the experiment is reached or a mirror instability driven $\beta$ limit occurs \citep{kotelnikov2024}.
The $\mratio=\{20,41,64\}$ cases operate with NBI power $\{200,400,1000\} \unit{kW}$ respectively.
The CQL3D-m/Pleiades loop is run for the duration of a laboratory shot, to $20
\unit{ms}$ (which is $t=0$ for Hybrid-VPIC).
At each end of the CQL3D-m run,
all three cases have plasma $\beta \sim 0.60$.
The low $\mratio=20$ (high $\vec{B}$-field) case achieves the highest plasma
density $1$--$3 \times 10^{13} \unit{cm^{-3}}$ on axis in the central cell
(Figure~\ref{fig:overview1d}(a)).
The ions have
$T_\mt{i} = \{13, 11, 11\} \unit{keV}$
at the origin $(r,z)=(0,0)$
in the $\mratio=\{20,41,64\}$ cases respectively.
Of note, the $\mratio=64$ case has a cooler ion plasma temperature
$T_\mt{i} \sim 5 \unit{keV}$ at the plasma's radial edge,
whereas the lower $\mratio$ (higher field) CQL3D-m simulations maintain
$T_\mt{i} \sim 10 \unit{keV}$ from the axis $r=0$ to the edge.
This is a result of the larger cool thermal ion population that is trapped by the sloshing-ion distribution in the $\mratio=64$ case.

The CQL3D-m bounce-averaged distribution function at the mirror's mid-plane ($z=0$) is
mapped on Liouville characteristics to all ($r,z$) and read into Hybrid-VPIC as
an initial condition for both real- and velocity-space ion distributions.
The CQL3D-m ion radial density profile $n$ is extrapolated
from $\sqrt{\psi_n} = 0.9$ to $1$ as
\begin{equation}
    n(\psi_n) = \cos^2 \left[
        \frac{\pi}{2}
        \left( \frac{\psi_n - 0.81}{1 - 0.81} \right)
    \right]
    \, .
\end{equation}
where $\psi_n = \psi / \psi_\mt{limiter}$,
$\psi = \int 2\pi B r\dtl r$,
and
$\psi_\mt{limiter} = 2.32 \times 10^6 \unit{G\, cm^2}$.
This sets the plasma's initial extent.
No limiter boundary condition is implemented in the Hybrid-VPIC simulation.

Electron velocity distributions
and the electrostatic potential $\phi$
are also solved in CQL3D-m via an
iterative technique \citep{frank2024}, but neither are directly input to Hybrid-VPIC's more-approximate fluid electron model.
Instead, we set the Hybrid-VPIC electron temperature $\Te = \{1.25, 1.5, 1.0\} \unit{keV}$
in the $\mratio=\{20,41,64\}$ cases respectively,
with $\Te$ values taken from the CQL3D-m simulation at $(r,z)=(0,0)$.
All simulations use an isothermal equation of state, so $\Te$ is constant
in space and time.
To support our use of a fluid approximation, we note that
the electron-electron collision time is much shorter than a WHAM shot duration,
so the CQL3D-m electron distributions are Maxwellians with empty loss-cones
beyond $v \sim \sqrt{e\phi/\me}$ (since the axial ambipolar potential confines
``core'' thermal electrons).
The overall $\Te$ varies by less than $2\times$ in both axial and radial
directions, within sloshing ion turning points, in the $\mratio=20$ case.

We use $N_\mt{ppc} = 8000$ ion macroparticles per cell, pinned to a reference
density $3 \times 10^{13} \unit{cm^{-3}}$, so the initial number of particles
is highest at the beam-ion turning points and lower elsewhere;
all particles have equal weight (or charge) in the PIC algorithm.

We initialize particles on their gyro-orbits with random gyrophase; this
spatially smooths the initial radial distribution of plasma density and
pressure, as compared to the CQL3D-m density distribution which places
particles at their gyrocenters.
The initial plasma in Hybrid-VPIC thus has non-zero initial azimuthal
diamagnetic drift and hence net angular momentum.
We also initialize the diamagnetic field from Pleiades in the Hybrid-VPIC
simulation, but our initial plasma and magnetic pressures are not in
equilibrium due to the Larmor radius offsets from particle gyrocenters.
Thus, the Hybrid-VPIC simulation evolves towards a new pressure equilibrium as
the plasma settles into steady state.

Finally, the initial electric field $\vec{E}(t=0)$ in Hybrid-VPIC
is given by Equation~\eqref{eq:ohm}
combined with the initial ion distributions from CQL3D-m,
our chosen values of $\Te$,
and the summed vacuum and diamagnetic $\vec{B}$ fields from Pleiades.

Let us define thermal length and time
normalizations.
The angular ion cyclotron frequency $\Omcio = e B(z=0)/(\mi c)$
at the mirror mid-plane.
The ion bounce (or, axial-crossing) time
$\tbounce = L_\mt{p} / \vtio \approx 1 \unit{\mu s}$
using the mirror's half length $L_\mt{p} = 98 \unit{cm}$ and a reference ion
thermal velocity
$\vtio = \sqrt{\Tio/\mi} = 0.00327 c = 9.8 \times 10^7 \unit{cm/s}$,
with $\Tio=20\unit{keV}$ and $c$ the speed of light.
Though the CQL3D-m initialized ions have $\Ti \sim 10\unit{keV}$,
our chosen $\vtio$ approximates
$\mi \vtio^2/2
\sim \mi v_\perp^2/2 \sim \mi v_\parallel^2/2
\sim (25 \unit{keV})/2$
for the beam-ion distribution's primary and secondary peaks.
We also define a reference ion Larmor radius
$\rLio = \vtio / \Omcio$ at the mirror mid-plane.
Tables~\ref{tab:physics} and \ref{tab:numerics} summarize physical and
numerical parameters, respectively, for our three fiducial simulations.

\begin{table*}
\centering
\begin{tabular}{@{}llllll p{4em}p{4em}p{4em}p{4em}@{}}
\toprule
    $\mratio$
    & $L_\mt{p}$
    & $B(z=0)$
    & $f_\mt{ci0}$
    & $\vtio$
    & $\rLio$
    & $T_\mt{e}$
    & Core $T_\mt{i}$ \newline at $0 \unit{\mu s}$
    & Edge $T_{\mt{i}\perp}$ \newline at $6\unit{\mu s}$
    & Edge $T_{\mt{i}\prll}$ \newline at $6\unit{\mu s}$ \\
\midrule
    20
    & $98 \unit{cm}$
    & $0.86 \unit{T}$
    & $6.5\unit{MHz}$
    & $980\unit{km/s}$
    & $2.37\unit{cm}$
    & $1.25 \unit{keV}$
    & $13\unit{keV}$
    & $8.4\unit{keV}$
    & $17.1\unit{keV}$ \\
    41
    & $\cdots$
    & $0.41$
    & $3.1$
    & $\cdots$
    & $4.94$
    & $1.5$
    & $11$
    & $6.5$
    & $13.1$ \\
    64
    & $\cdots$
    & $0.27$
    & $2.0$
    & $\cdots$
    & $7.64$
    & $1.0$
    & $11$
    & $6.4$
    & $8.0$ \\
\bottomrule
\end{tabular}
\caption{
    Physical parameters for fiducial simulations,
    labeled by vacuum mirror ratio $\mratio$.
    The ion cyclotron frequency $f_\mt{ci0} = \Omcio/(2\pi)$
    and ion Larmor radius $\rLio = \vtio/\Omcio$.
    Ions are deuterons.
    Core $\Ti$ at $0\unit{\mu s}$ is measured at the origin $(r,z)=(0,0)$.
    \label{tab:physics}
}
\end{table*}

\begin{table*}
\centering
\begin{tabular}{@{} lll lll lll@{}}
\toprule
    $\mratio$
    & $L_x$
    & $L_z$
    & $\Delta x$
    & $\Delta z$
    & $\Delta t$
    & $N_\mathrm{sub}$
    & $\eta$
    & $\eta_\mt{H}$ \\
\midrule
    20
    & $39.2 \unit{cm}$
    & $294 \unit{cm}$
    & $0.41 \unit{cm}$
    & $0.77 \unit{cm}$
    & $7.3\times 10^{-11} \unit{s}$
    & $100$
    & $0 \unit{s}$
    & $2.75\times10^{-14} \unit{s\,cm^2}$ \\
    41
    & $58.8$
    & $\cdots$
    & $\cdots$
    & $\cdots$
    & $\cdots$
    & $250$
    & $\cdots$
    & $\cdots$ \\
    64
    & $78.4$
    & $\cdots$
    & $\cdots$
    & $\cdots$
    & $\cdots$
    & $1000$
    & $\cdots$
    & $\cdots$ \\
\bottomrule
\end{tabular}
\caption{
    Numerical parameters for fiducial simulations, labeled
    by vacuum mirror ratio $\mratio$.
    \label{tab:numerics}
}
\end{table*}

\begin{figure*}
    \centering
    \includegraphics[width=\textwidth]{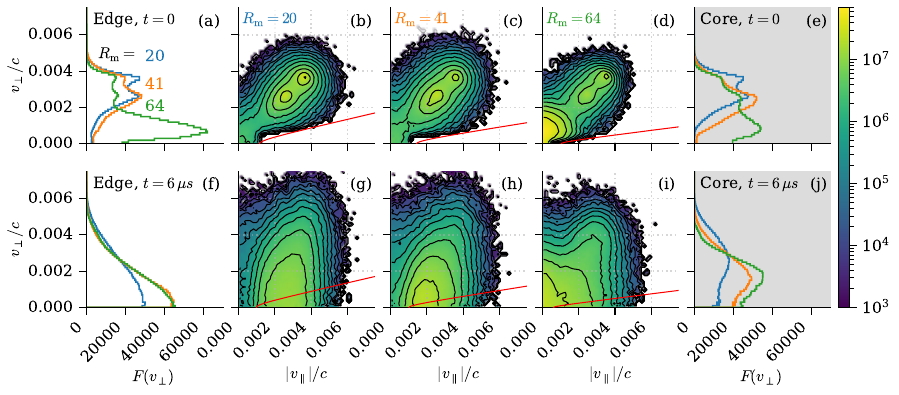}
    \caption{
        Initial (top row) and relaxed (bottom row) ion velocity distributions
        at the plasma edge, in three simulations.
        Edge ion distributions smooth and flatten in $v_\perp$ as the
        simulation evolves, with a stronger effect for edge plasma as compared
        to core plasma.
        The loss cone is filled, and the distribution varies little across the
        loss-cone boundary.
        (a): reduced distribution $F(v_\perp)$ for simulations with vacuum
        $\mratio = 20$ (blue), $41$ (orange), and $64$ (green).
        Distribution is normalized so that
        $\int F(v_\perp) 2\pi v_\perp \dtl v_\perp = 1$.
        (b)-(d): 2D distributions $f (v_\perp,v_\prll)$ for each of the three
        simulations shown in (a), normalized so that
        $\int f 2\pi v_\perp \dtl v_\perp \dtl v_\prll = 1$.
        Red curves plot loss-cone boundary, with the effect of electrostatic
        trapping approximated using the on-axis potential well depth of
        $0.4$ to $1.9 \unit{keV}$.
        (e): Like (a), but a ``core'' distribution centered on $r=0$ for
        comparison to the ``edge''.
        (f)-(j): like top row, but at later time $t=6\mu s$ in the simulation.
        In all panels, velocities $v_\perp$, $v_\prll$ are normalized to the
        speed of light $c$.
    }
    \label{fig:vdfs}
\end{figure*}

\section{Results}  \label{sec:results}

\subsection{Space, Velocity Structure of Steady-State Decay}
\label{sec:results-overview}

At the start of each simulation, the plasma relaxes from its initial state over
$\abt 1$--$3 \tbounce$; the diamagnetic field response is changed,
short-wavelength electrostatic fluctuations occur at the plasma edge, and
plasma escapes from the central cell into the expanders.
The plasma reaches a steady-state decay by $t = 6 \tbounce$ for all $\mratio$
simulations.  At this time,
(i) the particle loss time $\tloss = n / (\dtl n/\dtl t)$ is roughly constant
and exceeds the ion bounce time ($\tloss \gg \tbounce$),
(ii) the plasma beta
$\beta_\mathrm{i} = 8\pi P_\mathrm{i}/B^2 \sim 0.1$ to within a factor of two
at the origin $(r,z) = (0,0)$, with $P_\mathrm{i}$ the total ion pressure,
(iii) the combined vacuum and diamagnetic fields attain a mirror ratio
$\mratio = \{ 21, 45, 69 \}$ somewhat higher than the respective vacuum values
$\mratio = \{ 20, 41, 64 \}$.

Figure~\ref{fig:overview2d} shows the plasma's overall structure at
$t = 6 \tbounce$ for each of the vacuum $\mratio = 20, 41, 64$ simulations.
Flute-like, electrostatic fluctuations at the plasma's radial edge are visible
in $z=0$ slices of ion density and electric fields, with the strongest and most
coherent fluctuations for the $\mratio = 20$ case.
In the left-most panels (a), (e), (i), the axial outflow at
$|z| \gtrsim 70 \unit{cm}$ is split about $r = 0$, so more plasma escapes from
the radial edge $r > 0$ than the core $r \sim 0$.
In the right-most panels (d), (h), (l), the radial electric field fluctuation
$\delta E_r = E_r - \langle E_r \rangle_\theta$, where
$\langle \cdots \rangle_\theta$ represents an average over the azimuthal
coordinate to subtract the plasma's net radial potential.
The azimuthal fluctuation $\delta E_\theta$ in panels (c), (g), (k) is defined
similarly.
The transverse magnetic fluctuations $\delta B_r$ and $\delta B_\theta$ have
small amplitudes $\lesssim 10^{-3} B(z=0)$, whereas the electric fluctuations
$\delta E_r$ and $\delta E_\theta$ are of order $0.1 \vtio B(z=0)/c$,
corresponding to motional flows at thermal speeds.
We therefore neglect electromagnetic fluctuations and focus solely on the
azimuthal, electrostatic mode visible in Figure~\ref{fig:overview2d}.

Figure~\ref{fig:overview1d}(a)-(b) shows ion density profiles along $z$ both
on- and off-axis, with horizontal dashes marking the density floor imposed in
Ohm's Law, Equation~\eqref{eq:ohm}.
The off-axis density is measured along flux surfaces hosting the strong
electric fluctuations seen in Figure~\ref{fig:overview2d}.
Specifically, we pick surfaces at $r=\{9,18,22\}\unit{cm}$ and $z=0$
that have an approximate (azimuth-averaged, paraxial) flux coordinate
$\psi \approx \int 2\pi \langle B_z \rangle_\theta r \dtl r = \{ 2.1, 3.8, 3.7\} \times 10^6 \unit{G\,cm^{2}}$
for the $\mratio=\{20,41,64\}$ simulations respectively.
The plotted density $n_\mt{i}$ is also azimuth averaged.

The density profiles peak near the
turning points of $45^\circ$ pitch-angle ions,
defined as the $z$ locations where
the local mirror ratio $\mratio(z)=2$ on axis;
i.e. $\{36,44,57\} \unit{cm}$
(Figure~\ref{fig:overview1d}(c)).
Comparing on- and off-axis density peaks,
the off-axis peak is wider and decreases more slowly towards the mirror throat and expander.
This can be explained by the plasma edge's stronger loss-cone outflow and
broader pitch-angle distribution between $0^\circ$ and $45^\circ$, compared to
the plasma core at $r = 0$ (Figure~\ref{fig:vdfs}).

Figure~\ref{fig:overview1d}(d) shows on- and off-axis electrostatic potential
profiles $e\phi(z)/\Te$.
The off-axis profiles $\phi(s(z)) = -\int E_\prll(s) \dtl s$, with arc length
$s$ in the $(r,z)$ plane, are integrated along the same flux surfaces used in
Figure~\ref{fig:overview1d}(b).
We notice that the $z=0$ potential well has similar depth both on- and off-axis.
The density floor truncates the axial electrostatic potential at
$z \approx 100$ to $110 \unit{cm}$, so the full potential drop from the mirror
throat to the domain's $z$ boundary is not captured in our simulation.
In any case, plasma outflow in the expanders is not well modeled by our electron closure,
as the outflow is far from thermal equilibrium \citep[e.g.,][]{wetherton2021}.
We will restrict our attention to central-cell plasma behavior that we suppose to be
unaffected by the expanders.

\begin{figure*}
    \centering
    \includegraphics[width=0.8\textwidth]{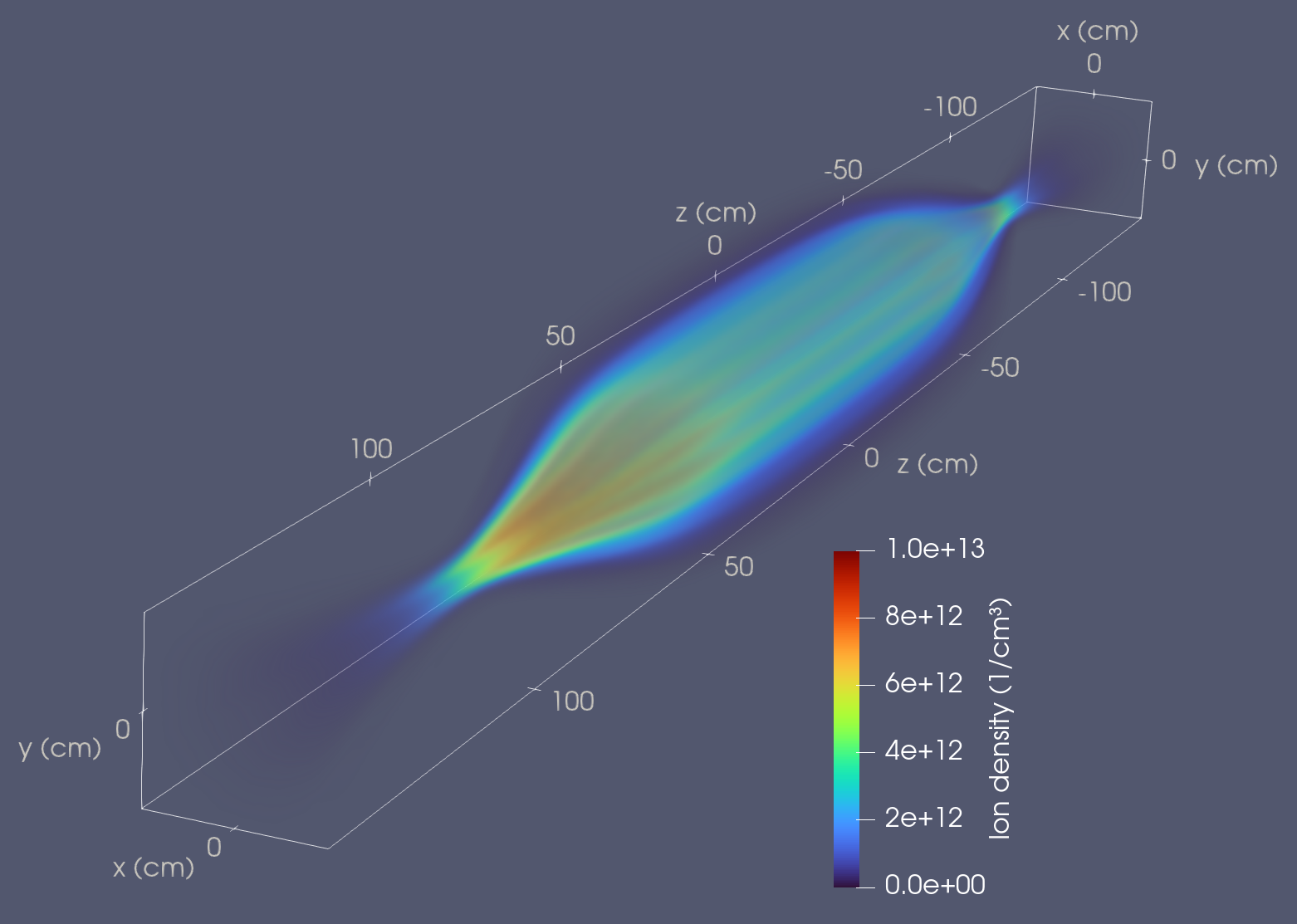}
    \caption{
        3D rendering of ion density in $\mratio=20$ simulation
        at $t=6\unit{\mu s}$;
        colormap is ion density in units of $\mathrm{cm}^{-3}$.
        An animated movie is available in the online journal.
    }
    \label{fig:movie}
\end{figure*}

Figure~\ref{fig:vdfs} shows initial ion velocity distributions, as imported
into Hybrid-VPIC from CQL3D-m, at the center of the mirror cell:
$z \in (-5.9, 5.9) \unit{cm}$ for all simulations.
Panels (a)--(d) sample ions from the plasma's radial edge:
$r \in [ 5.9, 11.8) \unit{cm}$ for $\mratio = 20$;
$r \in [11.8, 23.5) \unit{cm}$ for $\mratio = 41$;
$r \in [14.7, 29.4) \unit{cm}$ for $\mratio = 64$.
Panel (e) samples ions from the plasma's core:
$r \in [0,2.9) \unit{cm}$ for $\mratio = 20$;
$r \in [0,5.9) \unit{cm}$ for $\mratio = 41$;
$r \in [0,7.4) \unit{cm}$ for $\mratio = 64$.
Panels (f)-(j) shows ion distributions, selected from the same axial and radial
regions as the top row,
after the simulation has reached $t = 6 \tbounce \approx 6 \unit{\mu s}$.
Ions diffuse mostly in $v_\perp$; their distribution is continuous and nearly
flat across the velocity-space loss-cone boundary.
The reduced distribution $F(v_\perp) = \int f \dtl v_\parallel$ has
relaxed to a monotonically decreasing shape, $\dtl F / \dtl v_\perp < 0$, at the plasma edge
(Figure~\ref{fig:vdfs}(f));
however, the core plasma maintains $\dtl F / \dtl v_\perp > 0$ at low $v_\perp$
(Figure~\ref{fig:vdfs}(j)).
Some distribution function moments will be used in later discussion.
We define $\vec{B}$-perpendicular and parallel temperatures
$T_\mt{i\perp} \equiv (1/2) \int m_\mt{i} v_\perp^2 f \dtl\vec{v}$
and $T_\mt{i\prll} \equiv \int m_\mt{i} v_\prll^2 f \dtl\vec{v}$
so that $T_\mt{i} = (2T_\mt{i\perp}+T_\mt{i\prll})/3$;
temperature values for the edge ion distributions at $t=6\,\tbounce$ (Figure~\ref{fig:vdfs}(f)-(i))
are given in Table~\ref{tab:physics}.

Figure~\ref{fig:movie} shows a 3D render of ion density in the $\mratio=20$
simulation at $t=6\unit{\mu s}$.
The flute-like ($k_\parallel \sim 0$) nature of the edge fluctuations is
apparent.
An accompanying movie of the full time evolution from $t=0$ to $6\unit{\mu s}$
is available in the online journal.

To summarize, Figures~\ref{fig:overview2d}--\ref{fig:movie} show that at the
plasma's radial edge,
(i) flute-like electrostatic fluctuations appear,
(ii) axial outflow and hence losses are enhanced relative to the
plasma's core at $r \sim 0$,
and (iii) ions diffuse in $v_\perp$ to drive $\dtl F/\dtl v_\perp < 0$.
It is already natural to suspect that the electrostatic fluctuations diffuse
ions into the loss cone and hence cause plasma to escape the mirror.

\begin{figure*}
    \centering
    \includegraphics[width=\textwidth]{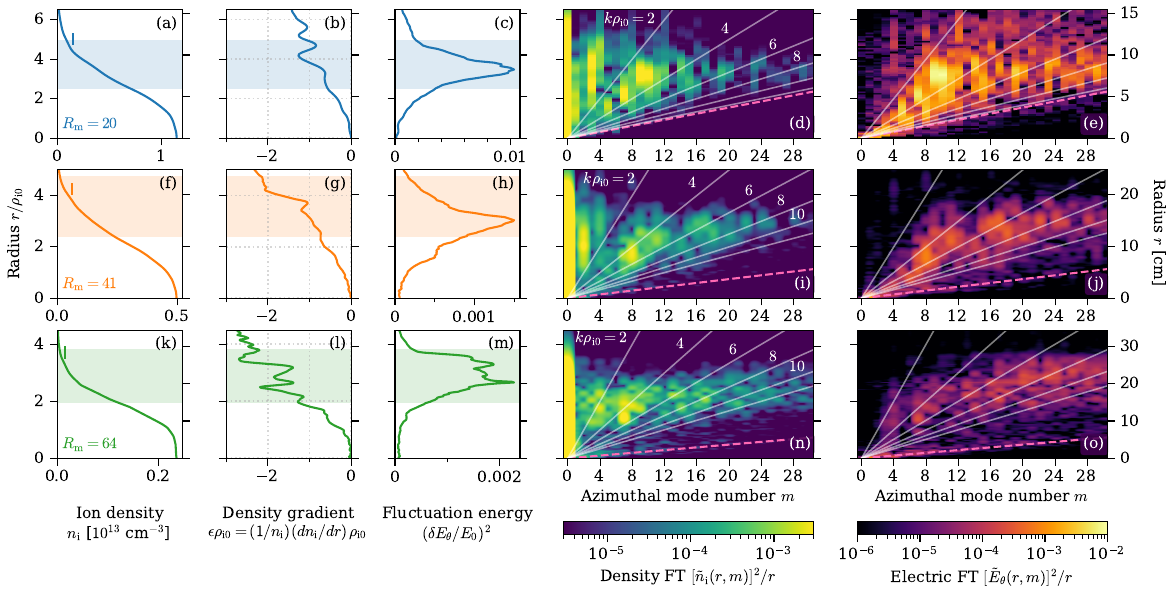}
    \caption{
        Radial structure of plasma at mid-plane $z=0$ and at
        $t = 6 \,\tbounce \approx 6 \unit{\mu s}$,
        for simulations with vacuum $\mratio=20$ (top row), $41$ (middle row),
        $64$ (bottom row).
        Left three columns show azimuth-averaged radial profiles of
        (a) ion density $n_\mt{i}$,
        (b) ion density gradient $\epsilon \rLio$,
        (c) azimuthal electrostatic fluctuation energy $\delta E_\theta^2$.
        Horizontal shaded bars
        contain the ``edge'' ion distributions from Figure~\ref{fig:vdfs}.
        Vertical dashes in left-most column mark density floor for
        Equation~\eqref{eq:ohm}.
        Right two columns (d)-(e) show azimuthal Fourier spectra of density
        $\tilde{n}_\mt{i}(r,m)$ and azimuthal electric field
        $\tilde{E}_\theta(r,m)$; Fourier transform maps $\theta \to m$, but
        radius $r$ is not transformed.
        White rays mark azimuthal wavenumber $k \rLio = 2,4,6,8,10,12$,
        with $k = m/r$.
        Dashed pink ray is the maximum $k = \pi/\Delta r$ resolved by the
        spatial grid, taking $\Delta r = \sqrt{2} \Delta x$.
        Panels (f)-(j) and (k)-(o) are organized similarly.
    }
    \label{fig:mode-k}
\end{figure*}

\subsection{Drift Cyclotron Mode Identification} \label{sec:dclc-id}

To establish the electrostatic mode's nature, we need to know plasma properties
at the radial edge and the mode's wavenumber and frequency spectrum.

Figure~\ref{fig:mode-k} (left three columns) presents the radial structure of the ion density
$n_\mt{i}$, and the electrostatic fluctuation energy
$\delta E_\theta^2 = \langle E_\theta^2 \rangle_\theta - \langle E_\theta \rangle_\theta^2$,
at the mirror mid-plane $z=0$.
Figure~\ref{fig:mode-k} (right two columns) also presents Fourier spectra of
density $\tilde{n}_\mt{i}(m,r)$ and electric component $\tilde{E}_\theta(m,r)$
as a function of azimuthal mode number $m$ and radius $r$.
Beware that Fourier spectrum normalization
is arbitrary here and in all figures;
Fourier amplitudes may be compared between panels within one figure,
but not across distinct figures.

The density gradient $\epsilon \equiv (\dtl n_\mt{i}/\dtl r)/n_\mt{i}$,
in units of inverse ion Larmor radius $\rLio^{-1}$, is of order unity and increases with
$\mratio$ (Figure~\ref{fig:mode-k}(b),(g),(l));
equivalently, the plasma column radius is smaller in units of $\rLio$ for
larger $\mratio$,
despite the column's larger physical extent.

The mode spectra of $\tilde{n}$ and $\tilde{E}_\theta$ suggest a partial
decoupling of density and electric fluctuations
(Figure~\ref{fig:mode-k}, right two columns).
In all simulations, low $m \sim 2$--$4$ density fluctuations are not
accompanied by a strong $E_\theta$ signal
(Figure~\ref{fig:mode-k}, right two columns).
The $\mratio=20$ simulation shows a strong mode in both density and $E_\theta$
fluctuations at $m \approx 9$--$10$ and
equivalent angular wavenumber $k \rLio \approx 2$--$4$
(Figure~\ref{fig:mode-k}(d)-(e)).
We identify this Fourier signal with phase-coherent fluting at the same $m$
visible to the eye in Figure~\ref{fig:overview2d}(b),(c).
In contrast, the $\mratio=41,64$ simulations show a decoupling of density and
$E_\theta$ fluctuations.
The strongest density fluctuations reside at
$r \sim 1$--$2\rLio$, $m \sim 7$--$8$, and $k\rLio \approx 2$--$6$
(Figure~\ref{fig:mode-k}(i),(n)),
whereas the electrostatic fluctuations reside at larger $r \sim 2$--$4\rLio$,
$m \sim 15$--$30$, and $k \rLio \sim 4$--$12$
(Figure~\ref{fig:mode-k}(j),(o)).

The fluctuations have $k_\parallel \ll k_\perp$ and are thus flute-like, which
we checked by plotting $\vec{E}$ in approximate flux-surface coordinates (not
shown).
Electric-field fluctuations terminate at the mirror throats and do not extend
into the expanders;
fluctuations may be artificially
truncated by the density floor
in Equation~\eqref{eq:ohm}.

\begin{figure*}
    \centering
    \includegraphics[width=\textwidth]{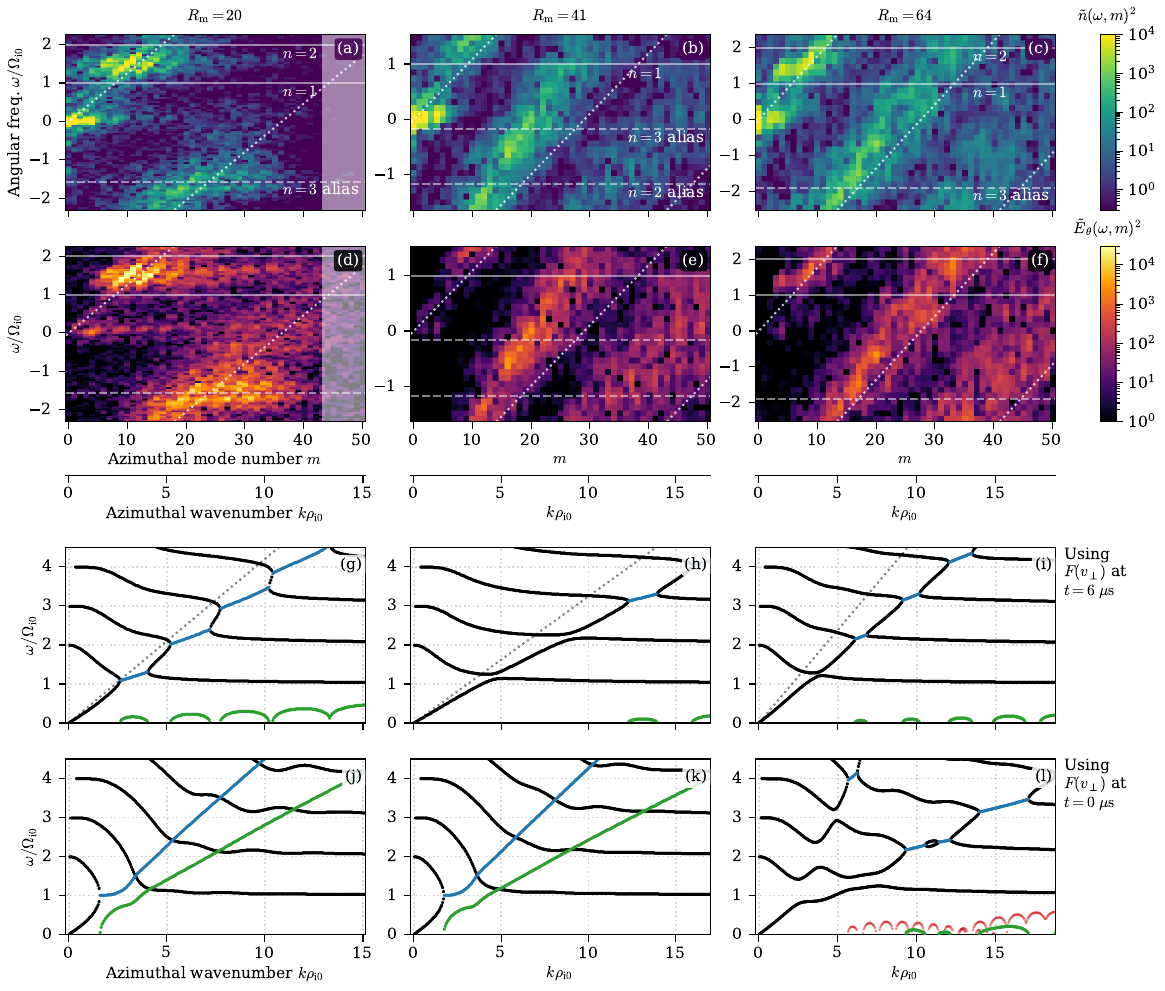}
    \caption{
        Time-azimuth Fourier spectra
        of density $\tilde{n}(\omega,m)^2$ (panels (a)-(c))
        and electric field $\tilde{E}_\theta(\omega,m)^2$ (panels (d)-(f))
        for simulations with $\mratio=\{20, 41, 64\}$ (left to right).
        Bottom rows show corresponding $(\omega,k)$ of unstable DCLC modes predicted by Equation~\eqref{eq:dclc-coldlec}
        for edge $F(v_\perp)$ at $t \approx 6 \unit{\mu s}$ (panels (g)-(i))
        or $t = 0$ (panels(j)-(l)).
        In panels (a)-(f),
        the full $\omega$ range within Nyquist-sampling limits is shown;
        signals with $\omega \gtrsim 2 \Omcio$ alias in frequency.
        White dotted lines plot ion diamagnetic drift velocity
        $\omega/k = v_\mt{Di}$.
        Shaded vertical bar in (a),(d) marks grid resolution
        limit $k > \pi/\Delta r$ with $\Delta r = \sqrt{2}\Delta x$.
        In panels (g)-(l), we plot
        both stable- and unstable-mode frequencies $\Real(\omega)$ (black, blue), and
        also the corresponding unstable-mode growth rates $\Imag(\omega)$ (green).
        In panel (l) only, red curves plot $\Imag(\omega)$ for higher-$\omega/k$
        modes with $\Real(\omega) \in [4\Omcio,14\Omcio]$ beyond the plot extent.
        Black dotted lines plot $\omega/k = v_\mt{Di}$.
    }
    \label{fig:mode-f}
\end{figure*}

Joint time-frequency and azimuthal-mode spectra of density and electric field
fluctuations, $\tilde{n}(\omega,m)$ and $\tilde{E}_\theta(\omega,m)$,
are presented in Figure~\ref{fig:mode-f}(a)-(f).
Fluctuations are sampled at radii $r = \{3.34, 2.98, 2.69 \} \rLio$
respectively, over $t = 3$ to $6 \,\tbounce$;
$\omega$ is angular frequency.
Positive $\omega/k$ corresponds to the ion diamagnetic drift direction.
We interpret Fourier power at $\omega < 0$ as high-$\omega$ signal that is
aliased in frequency space and would otherwise be contiguous in physical
$(\omega,k)$.
Assuming so, both $\tilde{n}$ and $\tilde{E}_\theta$ show a mode spectrum with
a phase speed $\omega/k > 0$ comparable to the ion diamagnetic drift speed
$v_\mt{Di}$ (white dotted lines, Figure~\ref{fig:mode-f}(a)-(f)).
We compute
\begin{equation}
    v_\mt{Di}
    \approx  \frac{T_\mt{i\perp}/m_\mt{i}}{\Omcio} \left(-\frac{1}{n_\mt{i}} \frac{\dtl n_\mt{i}}{\dtl r}\right)
    = \vtio \frac{T_\mt{i\perp}}{T_\mt{i0}} |\epsilon| \rLio
\end{equation}
using $\epsilon \rLio = \{ -1, -1, -1.5 \}$ and $T_\mt{i\perp}$ measured at
$t=6\tbounce$ (values reported in {\S}\ref{sec:results-overview}).
The spectra align with $\omega/k = v_\mt{Di}$ within a factor of $2$.

A fundamental mode appears at $\omega \in [\Omcio, 2 \Omcio]$ in all
simulations.
Fluctuation power extends to $\omega \gtrsim 3 \Omcio$ in all simulations,
perhaps up to $\omega \gtrsim 7 \Omcio$ in the $\mratio=64$ simulation, but by
eye we do not discern discrete harmonics above $2 \Omcio$.
A low-frequency $\omega \ll \Omcio$ mode with non-zero $m$ appears chiefly
in $\tilde{n}$ and weakly in $\tilde{E}_\theta$; we identify this slower motion
as fluid interchange and discuss it further in {\S}\ref{sec:other-modes}.

To help interpret
Figure~\ref{fig:mode-f}(a)-(f), we compute the linearly
unstable $(\omega,k)$ for DCLC in a planar-slab plasma with a spatial density
gradient $\epsilon$ and uniform background magnetic field ($\del B = 0$).
In such a plasma, a dispersion relation for exactly-perpendicular electrostatic
waves can be obtained by integrating over unperturbed orbits and Taylor
expanding $f$ in particle guiding-center coordinate, following
\citet[{\S}14-3, Eq.~(8)]{stix1992}.
The dispersion relation is then
\begin{equation} \label{eq:dclc-general}
    D = 1 + \sum_s \chi_s = 0
    \, ,
\end{equation}
where the perpendicular ($k=k_\perp$) susceptibility of species $s$ reads:
\begin{align} \label{eq:chi-general}
    \chi_\mt{s} = \left(\frac{\omps}{\Omcs}\right)^2 \bigg[
        \left( 1 - \frac{\epsilon \omega}{k} \right)
        \frac{1}{k^2}
        &\sum_{n=-\infty}^{\infty} \frac{n}{\omega - n}
        \int \dtl^3\vec{v}
            \left(
                \frac{1}{v_\perp}
                \ptlff{f}{v_\perp}
            \right)
            J_n^2 \nonumber \\
        - \frac{\epsilon}{k}
        &\sum_{n=-\infty}^{\infty} \frac{1}{\omega - n}
        \int \dtl^3\vec{v}
            f
            J_n^2
    \bigg]
    \, ,
\end{align}
In Equation~\eqref{eq:chi-general}, variables are written in a
\emph{species-specific} dimensionless form:
$\omega/\Omcs \to \omega$, $k\rLs \to k$, $\epsilon \rLs \to \epsilon$,
and $v_\perp/v_{\mathrm{t}s} \to v_\perp$,
where $\Omcs$ is signed (i.e., $\Omce < 0$)
and $\rLs \equiv v_{\mathrm{t}s}/\Omcs$.
The decision of how to define $v_{\mathrm{t}s}$ (with or without $\sqrt{2}$) is
given to the user.
The plasma frequency $\omps = \sqrt{4\pi n_\mt{s} q_\mt{s}^2/m_\mt{s}}$ for
each species.
The Bessel functions $J_n = J_n(k_\perp v_\perp)$ as usual, with $k_\perp=k$.
Equations~\eqref{eq:dclc-general}-\eqref{eq:chi-general} simplify for cold
fluid electrons to yield:
\begin{equation} \label{eq:dclc-coldlec}
    D
    = 1 + \chi_\mt{i}
    + \frac{\ompe^2}{\Omce^2}
    + \frac{\ompe^2}{|\Omce|} \frac{\epsilon}{k \omega}
    = 0
    \, ,
\end{equation}
the variables $k$, $\epsilon$, and $\omega$
are now in dimension-ful units.
Equation~\eqref{eq:dclc-coldlec} is the slab DCLC dispersion relation also used
by \citet[Eq.~(2)]{lindgren1976}, \citet[Eq.~(19)]{ferraro1987}, and
\citet[Eqs.~(17) and (A14)]{kotelnikov2017}.
In our sign convention, $\epsilon < 0$ obtains DCLC with $\omega/k > 0$ in the
ion diamagnetic drift direction.
Equation~\eqref{eq:dclc-coldlec} also hosts normal modes with $k < 0$ and high
phase velocity in the electron diamagnetic drift direction
\citep[{\S}2.A.1.b]{lindgren1976}, which do not appear in our simulations
and so are omitted from our discussion.

The unstable- and normal-mode solutions to Equation~\eqref{eq:dclc-coldlec},
presented in Figure~\ref{fig:mode-f}(g)-(l), are computed as follows.
First, we take $\Omcio$, $\rLio$, and $\vtio$ as defined in
{\S}\ref{sec:methods-plasma} to normalize all variables in
Equation~\eqref{eq:dclc-coldlec}.
Plasma parameters used for the $\mratio = \{ 20,41,64 \}$ simulations
respectively, are:
$\epsilon \rLio = \{ -1, -1, -1.5 \}$;
$n_\mt{i} = \{ 4, 1.2, 0.5 \} \times 10^{12} \unit{cm^{-3}}$;
and $B = \{ 8.6, 4.1, 2.7 \} \times 10^{3} \unit{G}$.
Both $\epsilon$ and $n_\mt{i}$ describe the plasma edge at the mid-plane $z=0$
(Figure~\ref{fig:mode-k}).
We take $B$ at $(r,z)=(0,0)$ to match the variable normalization throughout
this manuscript; $B$ at the plasma edge differs by $\lesssim 10\%$.
Reduced ion distributions $F(v_\perp) = \int f \,\dtl v_\prll$ are measured
directly from the plasma edge (Figure~\ref{fig:vdfs}).
Bessel function sums are computed using all terms with index $|n| \le 40$.
The waves and particles at hand have $k_\perp\rLio \lesssim 20$ and
$v_\perp/\vtio \lesssim 2$, so the Bessel function argument
$(k_\perp\rLio) (v_\perp/\vtio) \lesssim 40$.
Terms with $n > 40$ contribute little to $\chi_\mt{i}$ because the first
positive oscillation of $J_n(\xi)$ peaks at $\xi = j_n' > n$, where $j_n'$ is the
smallest positive zero of $J_n'$ \citep[{\S}15.3]{watson1922},
and $J_n(\xi) \to 0$ quickly as $\xi \to 0$ for $\xi \lesssim n$.

We then compute $D$ on a discrete mesh of $(k, \Real(\omega), \Imag(\omega))$; for
each $k$, we identify normal modes (whether stable, damped, or growing) by
seeking local minima of $D$ with respect to the complex $\omega$ mesh.
Our $D \approx 0$ solutions are not exact.
To test our solution scheme, we refined our solutions to $D=0$ by applying a
manual root-finder to each $(k,\omega)$ normal mode for one set of plasma
parameters, and we saw no significant difference.

Figure~\ref{fig:mode-f}(g)-(i) uses $F(v_\perp)$ measured from the plasma edge
at $t = 6 \tbounce \approx 6 \unit{\mu s}$, showing DCLC modes at marginal
instability (more precisely, drift-cyclotron modes since the loss cone is
filled).

Figure~\ref{fig:mode-f}(j)-(l) uses $F(v_\perp)$ measured at $t=0$ instead to
show that initial distributions with empty loss-cones and spatial gradient
$\epsilon\rLio \sim \mathcal{O}(1)$ drive strongly unstable, broad-band
electrostatic modes with fastest growth towards high $k\rLio \gg 1$ and
$\omega \gg \Omcio$.
The $\mratio=64$ simulation (Figure~\ref{fig:mode-f}(l)) is an exception,
because its CQL3D-m model predicts a larger population of trapped cool ions
that helps stabilize DCLC.
Figure~\ref{fig:mode-f}(l) also reveals three branches of unstable modes, each
with distinct $\omega/k$, that we speculate may be drift waves associated
with distinct hot and cool plasma populations (Figure~\ref{fig:vdfs}(a),(d)).
The slowest branch is visible with $\Real(\omega)$ between $2$ to $4\Omcio$;
the corresponding $\Imag(\omega)$ are plotted in green.
The faster phase speed branches have unstable $\Real(\omega) > 4 \Omcio$
extending to at least $14\Omcio$;
the corresponding $\Imag(\omega)$ are plotted in light red.

What is learned from comparing the simulation spectra versus linear theory in
Figure~\ref{fig:mode-f}?
First, marginally-stable DCLC mode growth may explain high $k\rLio \gtrsim 5$
fluctuations residing in the device during steady-state decay.
How do we explain the fundamental mode between $\omega=\Omcio$ and $2\Omcio$
for simulations with $\mratio=41,64$, since that mode is predicted to be
linearly stable at late times?
It may be an initially excited mode that did not damp and so persists to late
times; this appears possible for the $\mratio=41$ simulation, where the
fundamental is unstable at $t=0$.
Or, it may be excited by non-linear flow of wave energy from unstable to stable
modes; such an explanation may be needed for the $\mratio=64$ simulation, in
which cool plasma at the radial edge should quench DCLC growth of the
fundamental mode at both $t=0$ and $t=6\unit{\mu s}$.
We have interpreted the $t=0$ and $t=6\unit{\mu s}$ as most- and least-unstable
scenarios for DCLC growth, but the plasma may also transition through other
states that destabilize the fundamental mode.

\subsection{Ion Scattering} \label{sec:scattering}

\begin{figure}
    \centering
    \includegraphics[width=0.8\textwidth]{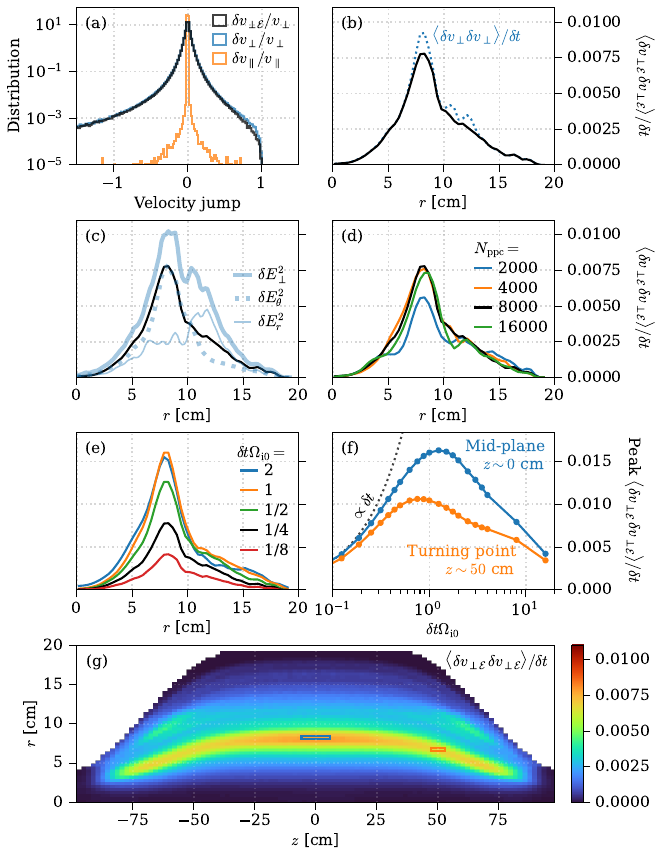}
    \caption{
        Ion scattering measured
        in $\mratio=20$ simulation,
        at mid-plane $z\in [-5.9,5.9] \unit{cm}$
        unless said otherwise.
        All diffusion coefficients are normalized to $\vtio^2\Omcio$.
        (a): Probability distribution of ion velocity jumps,
        normalized to $v_\perp(t_1)$ and $v_\prll(t_1)$, for particles at all radii.
        (b): Radial profile of
        ion diffusion $\langle \delta v_{\perp\mathcal{E}} \delta v_{\perp\mathcal{E}} \rangle/ \delta t$ (solid black)
        compared to $\langle \delta v_\perp \delta v_\perp \rangle/ \delta t$ (dotted blue).
        (c): Predicted radial profile of ion diffusion due to fluctuating fields
        $\delta E_\theta^2$ (dotted blue), $\delta E_r^2$ (thin solid blue),
        and $\delta E_\perp^2 = \delta E_\theta^2 + \delta E_r^2$ (thick solid
        blue), compared to
        $\langle \delta v_{\perp\mathcal{E}} \delta v_{\perp\mathcal{E}} \rangle/ \delta t$ (black).
        (d): Numerical convergence in particles per cell for
        radial profile of
        $\langle \delta v_{\perp\mathcal{E}} \delta v_{\perp\mathcal{E}} \rangle/ \delta t$.
        (e): Effect of
        measurement time $\delta t$ upon radial
        profile of $\langle \delta v_{\perp\mathcal{E}} \delta v_{\perp\mathcal{E}} \rangle/ \delta t$.
        (f): Effect of
        measurement time $\delta t$ upon diffusion
        measured at the mid-plane $(r,z) \approx (8.2, 0) \unit{cm}$ (blue curve),
        and near the beam-ion turning point at $(r,z) \approx (6.8, 50) \unit{cm}$ (orange curve).
        (g) 2D map of diffusion
        $\langle \delta v_{\perp\mathcal{E}} \delta v_{\perp\mathcal{E}} \rangle/ \delta t$
        computed in discrete $(r,z)$ bins (pixels); only bins with $> 100$
        particles are shown.
        Light blue and orange boxes mark measurement locations used in panel (f).
    }
    \label{fig:jump-moments}
\end{figure}

To establish a causal link between $\delta E$ fluctuations and axial ion
losses, we quantify ion scattering in the $\mratio=20$ simulation as follows.
We measure velocity jumps over a short time interval
$\delta t \equiv t_1 - t_0 = 0.25 \Omcio^{-1}$ for $\mathcal{O}(10^7)$ PIC
macro-particles sampled from $|z| \in [0, 5.9] \unit{cm}$.
Our approach is similar to
many
other PIC simulation studies; see
\citet{yerger2025} for a recent discussion of nuances in constructing and
interpreting such velocity jump moments.
Figure~\ref{fig:jump-moments}(a) shows the probability distribution of the
$\vec{B}$-perpendicular and parallel velocity jumps,
$\delta v_\perp = v_\perp(t_1) - v_\perp(t_0)$
and $\delta v_\prll = v_\prll(t_1) - v_\prll(t_0)$.
The distributions are not Gaussian and have long tails.
The perpendicular jumps $\delta v_\perp$ are much larger than $\delta v_\prll$,
as expected for flute-like ($k_\perp \gg k_\prll$) electrostatic modes and as
evident in Figure~\ref{fig:vdfs}.

Ion velocities may jump due to both adiabatic and non-adiabatic motion.
To separate these motions, introduce an energy
$\mathcal{E} = m_\mt{i} v^2/2 + e \langle\phi\rangle_{\theta,t}$,
where $\langle\cdots\rangle_{\theta,t}$ is an average over both azimuth angle
$\theta$ and time from $t_0$ to $t_1$.
We expect $\mathcal{E}$ to be conserved by particles gyrating in slowly-varying
$\vec{E}$ and $\vec{B}$ fields, at lowest order in a Larmor-radius expansion.
Therefore, we attribute jumps in $\mathcal{E}$ to a non-adiabatic kick in
perpendicular velocity that we call $\delta v_{\perp \mathcal{E}}$.
We use
$\delta \mathcal{E} = \mathcal{E}(t_1) - \mathcal{E}(t_0) = m v_\perp \delta v_{\perp \mathcal{E}} - m (\delta v_{\perp \mathcal{E}})^2/2$
to compute:
\begin{equation} \label{eq:dvperpe}
    \frac{\delta v_{\perp \mathcal{E}}}{v_\perp(t_1)}
    = 1 - \sqrt{1 - \frac{2 \delta \mathcal{E}}{m v_\perp^2(t_1)}}
    \, .
\end{equation}
Equation~\eqref{eq:dvperpe} requires $\delta \mathcal{E}$ to not exceed the
particle's final perpendicular energy:
\begin{equation} \label{eq:dvperpe-select}
    \delta \mathcal{E} < m v_\perp^2(t_1)/2
    \, .
\end{equation}
Figure~\ref{fig:jump-moments}(a) shows that the probability distribution of
$\delta v_{\perp \mathcal{E}}$, computed only for those particles satisfying
Equation~\eqref{eq:dvperpe-select}, is marginally narrower than that of
$\delta v_\perp$, as expected if non-adiabatic kicks are the main contribution
to $\delta v_\perp$.\footnote{
    We checked that Equation~\eqref{eq:dvperpe-select} does not cause
    noticeable selection bias for short $\delta t \lesssim \Omcio^{-1}$;
    radial profiles of $\langle \delta v_\perp \rangle$ and
    $\langle \delta v_\perp \delta v_\perp \rangle$, computed with and without
    particles excluded by Equation~\eqref{eq:dvperpe-select}, appear identical to
    the eye.
    For larger $\delta t$, particles accumulate order-unity kicks in $v_\perp$
    and selection bias appears.
}

The ion diffusion $\langle \delta v_{\perp\mathcal{E}} \delta v_{\perp\mathcal{E}} \rangle/\delta t$
as a function of radius is shown in Figure~\ref{fig:jump-moments}(b);
its value is normalized to
$\vtio^2 \Omcio$
in all of
Figure~\ref{fig:jump-moments}(b)-(g).
Here, $\langle\cdots\rangle$ is a velocity-distribution moment computed in
radial bins.
The use of $\delta v_{\perp\mathcal{E}}$ decreases the measured diffusion as
compared to $\langle \delta v_\perp \delta v_\perp \rangle/\delta t$, as expected.

The ion diffusion due to fluctuating fields $\delta E_\perp(\vec{r})$ can be
described by a diffusion coefficient similar to those used in quasi-linear
models:
\begin{equation} \label{eq:DQL}
    D_{\perp\perp} = \frac{1}{2} \left( \frac{e}{m_\mt{i}} \delta E_\perp \right)^2 \tau_\mt{c}
    \, ,
\end{equation}
where $\tau_\mt{c}$ is a yet-unknown wave-particle correlation time.
Equation~\eqref{eq:DQL} assumes (i) weak but \emph{coherent} kicks
$\delta v_\perp \approx (e/m_\mt{i}) \delta E_\perp \tau_\mt{c}$;
(ii) a uniform random distribution of angles between $\vec{v}_\perp$ and
$\delta \vec{E}_\perp$ to obtain a factor of $1/2$ accounting for kicks in
gyrophase instead of $v_\perp$ magnitude.
For a scattering-measurement time $\delta t < \tau_\mt{c}$, we expect
\begin{equation} \label{eq:DQL-short-t}
    D_{\perp\perp}
    \approx
    \frac{
    \langle \delta v_{\perp\mathcal{E}} \, \delta v_{\perp\mathcal{E}} \rangle
    }{
        \delta t
    }
    \, ,
\end{equation}
also replacing $\tau_\mt{c} \to \delta t$ in $D_{\perp\perp}$.

Choosing $\delta t < \tau_\mt{c}$ is unusual for studies of particle
diffusion, as the resulting Equation~\eqref{eq:DQL-short-t} describes a more
``ballistic'' than diffusive process.
But, a short $\delta t$ helps us.
When using a longer $\delta t \gg \tau_\mt{c}$, at least two issues arise.
First, ions gyrate in and out of the scattering zone, as the zone's radial
width is similar to an ion Larmor radius.
A typical ion may get one or a few kicks, gyrate out of the scattering zone and
drift adiabatically, re-enter the scattering zone to be kicked again, and so
on, resulting in a random walk with intermittent large time gaps.
The scattering zone's finite radial width may also introduce bias in the
correlation time $\tau_\mt{c}$, because
a typical inboard (small $r$) ion gyrating in and out of the scattering zone
sees a redshift $\omega - k_\perp v_\perp$,
whereas a typical outboard (large $r$) ion instead sees a blueshift
$\omega + k_\perp v_\perp$.
Second, a longer $\delta t$ needed to sample multiple gyration periods
$2\pi\Omcio^{-1}$ will introduce axial bounce effects.
In the $\mratio=20$ simulation, $\tbounce \approx 40 \Omcio^{-1}$, and even
fewer ion gyrations are executed within $\tbounce$ for the higher $\mratio$
cases.

In Figure~\ref{fig:jump-moments}(c) we compare Equation~\eqref{eq:DQL-short-t}
to the ion diffusion measured from individual particles.
The fluctuating energy density is azimuth averaged as
$\delta E_r^2 = \langle E_r^2 \rangle_\theta - {\langle E_r \rangle_\theta}^2$, and
similarly for $\delta E_\theta^2$;
the sum $\delta E_\perp^2 = \delta E_\theta^2 + \delta E_r^2$.
The diffusion due to $\delta E_\theta^2$ agrees especially well with the particle
measurement, whereas the diffusion due to $\delta E_r^2$ agrees less well.

Numerical noise might drive axial losses from the plasma edge in the same way
that we are attributing to DCLC, because PIC particle count decreases at the
plasma edge.
To check this possibility, Figure~\ref{fig:jump-moments}(d) shows that the
measured ion scattering is converged in the number of particles per cell used.
We are confident that ion scattering is not due to numerical noise because
(i) the DCLC electric fields have much larger energy density than numerical
noise at the grid scale, and Figure~\ref{fig:jump-moments}(c) shows good
agreement in radial profiles of electric fields and scattering,
(ii) we see weak to no $N_\mt{ppc}$ dependence of scattering rates, whereas if
scattering were due to noise, we might expect either an outwards shift in $r$
as $N_\mt{ppc}$ increases (for fixed DCLC amplitude), or a decrease in
scattering rate if noise suppresses DCLC amplitude;
(iii) ion scattering is clearly anisotropic (Figure~\ref{fig:jump-moments}(a)),
whereas numerical scattering should be insensitive to $v_\parallel$ versus
$v_\perp$ because the grid scale is much smaller than the ion Larmor radius.

In Figure~\ref{fig:jump-moments}(e) we show the effect of $\delta t$ upon the
radial profiles of measured ion diffusion.
Figure~\ref{fig:jump-moments}(f) then samples the ion scattering at its radial
peak $r = 8.2 \unit{cm}$ (blue curve) and shows its dependence upon many more
values of $\delta t$.
We see that the diffusion moment scales linearly with small $\delta t$ as
expected from Equation~\eqref{eq:DQL-short-t}; for comparison, the black dotted
line shows an exactly linear correlation with $\delta t$.
As $\delta t$ becomes $\gtrsim \Omcio^{-1}$, waves and particles decorrelate
and the diffusion rate begins to fall.
We perform a similar calculation at $(r,z) \approx (6.8,50) \unit{cm}$
(Figure~\ref{fig:jump-moments}(f), orange curve) to conclude that $\tau_\mt{c}$ is
shorter near the fast ion turning point.
If $\tau_\mt{c} \sim 1/\Omci$
(where $\Omci$ varies with $z$, unlike $\Omcio$),
the lower $\tau_\mt{c}$ can be easily explained by
the $2\times$ increase in $B$ magnitude.

Finally, Figure~\ref{fig:jump-moments}(g) shows
$\langle \delta v_{\perp\mathcal{E}} \delta v_{\perp\mathcal{E}} \rangle/\delta t$
as a function of $(r,z)$ in the mirror's central cell.
The ion scattering at all $z$ is well localized to the same flux surfaces
between beam-ion turning points.
Scattering is strongest towards $z=0$, where the central-cell field is
relatively uniform.

We conclude from Figure~\ref{fig:jump-moments}(e)-(f) that particle scattering
has longer correlation time $\tau_\mt{c}$ and reaches larger amplitude at the
mirror mid-plane $z=0$, as compared to near the beam-ion turning points.
Ions at $z=0$, and throughout the central cell where $B \approx B(z=0)$, should
be more important for regulating DCLC growth and saturation than ions at the
turning points.

\subsection{Particle Confinement Time} \label{sec:confinement}

\begin{figure}
    \centering
    \includegraphics[width=0.8\textwidth]{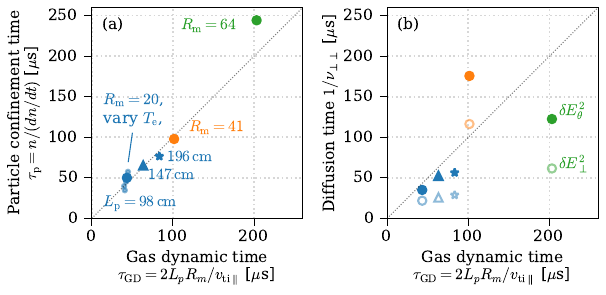}
    \caption{
        (a): Particle confinement time measured between $t=5$ to $6\tbounce$,
        for mirrors of varying $\mratio$ (blue, orange, green)
        and device length $L_\mt{p}$ (circle, triangle, star markers)
        as a function of $\tau_\mt{GD}$ (Equation~\eqref{eq:tau-gdt}).
        Small blue markers vary $\Te$ for $\mratio=20$; large blue marker is
        fiducial $\Te=1.25\unit{keV}$.
        (b): Diffusion timescale $1/\nu_{\perp\perp}$
        (Equation~\eqref{eq:nuQL}) modeled from $\delta E_\theta^2$ (solid
        markers) and $\delta E_\perp^2$ (hollow markers), as a function of
        $\tau_\mt{GD}$.
        In both panels, dotted black line is $\tau_\mt{p} = \tau_\mt{GD}$.
    }
    \label{fig:confinement-time}
\end{figure}

Because the loss cone is full---i.e., $F(v_\perp)$ is roughly constant within
the loss cone (Figure~\ref{fig:vdfs})---our simulated mirrors are a
collisionless analog of the Gas Dynamic Trap (GDT) at the Budker Institute
\citep{ivanov2017}.
Ions scatter across the loss-cone boundary as fast as (or faster than)
untrapped ions can stream out of the mirror, implying an effective mean free
path shorter than the device's length.
The particle confinement time
$\tau_\mt{p} \equiv N / |\dtl N/ \dtl t|$,
where $N$ is the total number of ions,
then scales like the eponymous ``gas dynamic'' time:
\begin{equation} \label{eq:tau-gdt}
    \tau_\mt{GD}
    = \frac{ 2 L_\mt{p} \mratio } { v_{\mt{ti}\prll} }
    \, ,
\end{equation}
adapted from \citet[{\S}3]{endrizzi2023} with $v_{\mt{ti}\prll}$ a characteristic
parallel thermal velocity.

To test the relation $\tau_\mt{p} \propto \tau_\mt{GD}$,
we measure $\tau_\mt{p}$ between $t = 5$ to $6\, \tbounce$, and
$v_{\mt{ti}\prll} = \langle v_\parallel^2 \rangle^{1/2}$ at
$t=6\,\tbounce$, in each of the $\mratio=\{20,41,64\}$ simulations with
$L_\mt{p} = 98 \unit{cm}$ on axis.
We also measure $\tau_\mt{p}$ and $v_{\mt{ti}\prll}$ in additional
$\mratio = 20$ simulations with varying $\Te = {0,2.5,5,10}$ keV and longer
central cells (larger $L_\mt{p}$); the latter are constructed as follows.
Split the ``original'' mirror device in half at $z=0$.
Between the mirror halves, insert a cylindrical plasma of length $98$ or $168$
cm, thereby increasing the entire mirror's half-length $L_\mt{p}$ by $1.5$ or
$2\times$.
The cylinder has, at all $z$, the same velocity distribution and magnetic field
$\vec{B}$ as in the original mirror at $z=0$.
The simulation domain is made larger; mesh voxel dimensions ($\Delta x$,
$\Delta y$, $\Delta z$) are the same as in {\S}\ref{sec:methods}.
The cylinder's magnetic field
is unphysical because it has $\dtl B_z/\dtl r \ne 0$ and $B_r = 0$,
implying non-zero current $c \del\times \vec{B}/(4\pi)$,
so we exclude this current from the $\vec{j} \times \vec{B}$ term in Ohm's Law
(Equation~\eqref{eq:ohm}).

The confinement time $\tau_\mt{p} \sim \mathcal{O}(10^2) \unit{\mu s}$,
and $\tau_\mt{p}$
scales linearly with $\tau_\mt{GD}$ as expected
(Figure~\ref{fig:confinement-time}(a)).
Gas-dynamic confinement explains losses from the $\mratio = 20$ and $41$
simulations very well.
Raising electron temperature $\Te$ from $0$ to $10 \unit{keV}$
lowers $\tau_\mt{p}$ from $57$ to $35 \unit{\mu s}$
for the $\mratio = 20$ simulations.
For comparison, the collisional (aka ``classical'') confinement time
is $0.1$--$0.2 \unit{s}$,
using Equation~(3.4) of \citet{endrizzi2023}
with $n=3\times10^{13}\unit{cm^{-3}}$,
$\mratio = 20$ to $64$, and beam energy $25\unit{keV}$.

The $\mratio = 64$ shows 20\% better particle confinement than predicted by
Equation~\eqref{eq:tau-gdt}.
Why?
The larger plasma radius and hence longer flux-tube length $> 2 L_\mt{p}$
between mirror throats only explains $\abt 5\%$ of the disagreement.
We speculate that electrostatic potential effects may explain the remaining
disagreement.
In the $\mratio =20,41$ cases, beam ions diffuse in $v_\perp$ and escape with
high $v_\prll$; electrostatic effects are weak since $\Te \ll \Ti$, so
Equation~\eqref{eq:tau-gdt} accurately describes the beam ion confinement.
The $\mratio = 64$ case has more cool, low-temperature ions
(Figure~\ref{fig:vdfs}(i)) that can be trapped by the sloshing-ions' potential
peak at $z \sim 30 \unit{cm}$
\citep{kesner1973,kesner1980};
confinement is thus improved.

Instabilities in many settings are self regulating; i.e., unstable waves drive
phase-space flow that quenches the waves' own energy source, driving the system
to equilibrium \citep[e.g.,][]{kennel1966-petschek}.
If DCLC self regulates, then we may expect its amplitude to grow in time until
the diffusion rate into the loss cone balances the axial outflow rate:
$\nu_{\perp\perp} \propto \tau_\mt{GD}^{-1}$.
We test this by computing a diffusion rate into the loss cone as:
\begin{equation} \label{eq:nuQL}
    \nu_{\perp\perp} \equiv
        \frac{1}{N_\mt{\ell}}
        \int
        \frac{ D_{\perp\perp}(r) }{ {v_{\perp\mt{LC}}}^2 }
        n_\mt{i}(r) 2\pi r \ell
        \,\dtl r
        \, ,
\end{equation}
which is a density-weighted average of $D_{\perp\perp}$
(Equation~\eqref{eq:DQL}) over a cylindrical kernel of axial length $\ell$
and radial profile $n_\mt{i}(r)$, normalized to
$N_\mt{\ell} = \int n_\mt{i}(r) 2\pi r \ell \dtl r$.
We take $v_{\perp\mt{LC}} = v_{\mt{ti}\prll}/\sqrt{\mratio - 1}$ to
approximate the ions' $v_\perp$ at the loss cone boundary,
we take $\ell = 12 \unit{cm}$ centered at $z=0$,
and we take $\tau_\mt{c} = \Omcio^{-1}$.
Given these assumptions, and given that Equation~\eqref{eq:DQL} is not from a
self-consistent quasi-linear theory, we interpret Equation~\eqref{eq:nuQL} as
no more accurate than an order-of-magnitude scaling.
In Figure~\ref{fig:confinement-time}(b) we compute the diffusion timescale
$1/\nu_{\perp\perp}$ using either $\delta E_\theta^2$ or $\delta E_\perp^2$ as
defined as in Figure~\ref{fig:jump-moments}(c).
We observe that $1/\nu_{\perp\perp}$ has similar magnitude as $\tau_\mt{GD}$,
as expected.
But, no trend is obvious from the scatter and few data points.

\section{Discussion} \label{sec:discussion}

\begin{figure}
    \centering
    \includegraphics[width=0.85\textwidth]{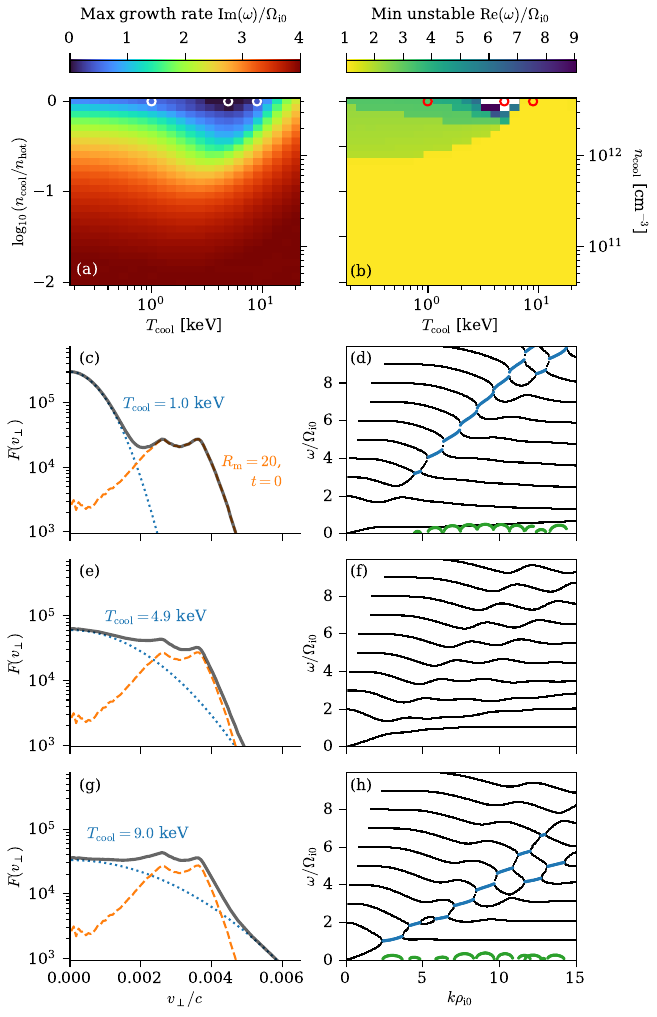}
    \caption{
        Effect of cool plasma on DCLC linear stability in WHAM with
        $\mratio=20$ and a hot beam-ion distribution.
        (a): 2D regime map of maximum growth rate $\Imag(\omega)/\Omcio$ as a
        function of $n_\mt{cool}$ and $T_\mt{cool}$.
        (b): Like (a), but showing minimum $\Real(\omega)/\Omcio$ that is
        DCLC unstable.  As cool plasma density is raised, low harmonics are
        stabilized.
        White pixels at top of panel ($T_\mt{cool} \sim 5 \unit{keV}$
        and $\log_{10}(n_\mt{cool}/n_\mt{hot}) \sim 0$)
        mean that no linearly-unstable modes were found.
        (c): Example ion distribution $F(v_\perp)$ with $1 \unit{keV}$ cool
        plasma (dotted blue) added to initial $\mratio=20$ distribution.
        (d): Dispersion relation solutions corresponding to (c), showing normal
        modes (black), unstable mode $\Real(\omega)$ (blue), and unstable mode
        $\Imag(\omega)$ (green).
        (e-f): like (c-d), but with $4.9 \unit{keV}$ cool plasma.
        (g-h): like (c-d), but with $9.0 \unit{keV}$ cool plasma.
    }
    \label{fig:cool-plasma}
\end{figure}

\begin{figure}
    \centering
    \includegraphics[width=\textwidth]{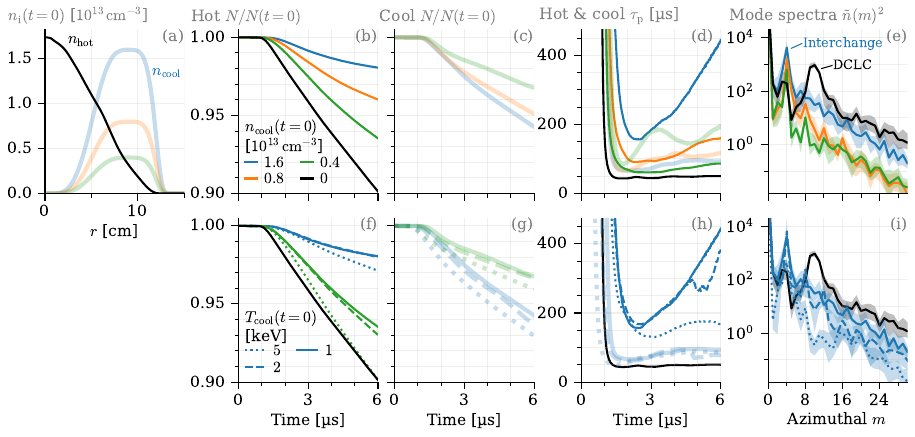}
    \caption{
        Effect of cool plasma on particle losses and density fluctuations
        in Hybrid-VPIC simulations with $\mratio=20$.
        (a): Initial density radial profiles for hot (black) and cool (colored) ions.
        (b): Total number of hot ions within simulation domain, normalized to
        initial value, for varying $n_\mt{cool}$ at fixed $T_\mt{cool}=1\unit{keV}$.
        (c): Like panel (b), but for cool ions.
        (d): Particle confinement time $\tau_\mt{p} = N/(\dtl N/\dtl t)$
        for hot and cool populations, same simulations as in panels (b)--(c).
        (e): Fourier spectra of azimuthal density fluctuations, using total
        (hot plus cool) ion populations; solid lines are median and shading is
        25--75 percentile range within $3$--$6\unit{\mu s}$.
        (f)--(i): like top row (b)--(e), but emphasis on
        varying $T_\mt{cool} = \{1,2,5\} \unit{keV}$ at fixed $n_\mt{cool}$.
        Curve styles are matched across all panels.
    }
    \label{fig:cool-confinement}
\end{figure}

\subsection{Cool Plasma Effects} \label{sec:cool}

How much cool plasma, and at what temperature, suppresses DCLC for the
peaked beam-ion distributions injected into WHAM?
To answer this, Figure~\ref{fig:cool-plasma} computes DCLC linear stability
with distinct ``hot'' and ``cool'' ion populations.
The hot ions are a beam distribution at $t=0$ in our $\mratio=20$ simulation,
taken from the mid-plane $z=0$ (Figure~\ref{fig:vdfs}(b)),
with $n_\mt{hot} = 4 \times 10^{12} \unit{cm^{-3}}$.
The cool ions are a Maxwellian of the same species (deuterium), with density
$n_\mt{cool}$ and temperature $T_\mt{cool}$.
We solve Equation~\eqref{eq:dclc-coldlec} using the same procedure as in
{\S}\ref{sec:dclc-id}, within a finite domain $k\rLio < 15$,
$\Real(\omega)/\Omcio < 10$, and $\Imag(\omega)/\Omcio < 4$.

Figure~\ref{fig:cool-plasma}(a) predicts that DCLC is suppressed when
cool and hot ion densities are nearly equal,
and $T_\mt{cool} \sim 2$ to $10 \unit{keV}$.
In cases where DCLC is not fully stabilized, panel (b) shows that dense-enough
cold plasma will at least stabilize low cyclotron harmonics; we anticipate that
the remaining unstable high harmonics may have weaker scattering rate.
The Figure~\ref{fig:cool-plasma}(b) prediction qualitatively concurs with
recent measurements on the GDT device: DCLC at high harmonics appeared when a
relatively high gas density was puffed into GDT's central chamber before
neutral-beam injection \citep{prikhodko2018,shmigelsky2024-dclc};
critically, this form of DCLC did not impede build-up of plasma pressure.
Panels (c)-(h) show the effect of varying $T_\mt{cool}$
(with $n_\mt{cool} = n_\mt{hot}$) upon DCLC mode structure in $(\omega,k)$.
As $T_\mt{cool}$ rises, quenching of low harmonics proceeds to total
stabilization.
When $T_\mt{cool}$ is too high and near the beam ions' effective temperature,
the ``cool'' plasma is less able to reduce the velocity-space gradient
$\dtl F/\dtl v_\perp$ and DCLC become unstable at all harmonics.

To test the predictions of Figure~\ref{fig:cool-plasma}, we repeat the
$\mratio=20$ Hybrid-VPIC simulation with cool plasma added to the radial edge,
varying $n_\mathrm{cool} \approx \{ 4, 8, 16 \} \times 10^{12} \unit{cm^{-3}}$
within radii $r \sim 5$ to $12 \unit{cm}$ (Figure~\ref{fig:cool-confinement}(a))
and also varying $T_\mt{cool} = \{ 1, 2, 5 \} \unit{keV}$.
The simulation results are summarized in Figure~\ref{fig:cool-confinement}(b)-(i).

Cool $1\unit{keV}$ plasma quenches DCLC losses and improves the hot plasma's
final $N/N(t=0)$ by a factor of $\abt 2$ to $5\times$
(Figure~\ref{fig:cool-confinement}(b)).
The hot plasma's final confinement time is at most
$\tau_\mt{p} = 445 \unit{\mu s}$, a stark improvement over
$\tau_\mt{p} = 50 \unit{\mu s}$ without cool plasma
(Figure~\ref{fig:cool-confinement}(d)).
The cool plasma itself is less well confined with
$\tau_\mt{p} \lesssim 200 \unit{\mu s}$ (Figure~\ref{fig:cool-confinement}(d)),
but it may be externally replenished in real experiments or (future) more
realistic future simulations.
The simulation with lowest $n_\mt{cool} = 4 \times 10^{12} \unit{cm^{-3}}$
shows that cool ions are better confined than the hot ions,
qualitatively consistent with trapping by the sloshing ions' axial potential
(Figure~\ref{fig:cool-confinement}(b)-(c)).
The situation reverses at higher $n_\mt{cool}$: hot ions become better confined
than cool ions, which we speculate may be due to flattening of ion density $n$,
and hence also electric potential $\phi$, along $z$.

Azimuthal fluctuations in density confirm that cool plasma quenches DCLC at
$m \approx 10$ (Figure~\ref{fig:cool-confinement}(e)).
But, cool ions also drive faster growing MHD interchange-like modes
at $m \approx 4$.
If our simulations were run longer than $t = 6 \unit{\mu s}$, these
interchanges might eventually cause large ion losses.
In laboratory devices, interchange can be stabilized by shear flow driven by
either external voltage biasing \citep{beklemishev2010,yakovlev2018}
or electron cyclotron heating \citep{yoshikawa2019}.
We thus remain optimistic that cool plasma stabilization can work in WHAM,
especially given the method's success in real laboratory experiments
\citep{coensgen1975,shmigelsky2024-dclc}.
In \S\ref{sec:other-modes}, we will comment further on interchange
identification and growth/suppression.

The Hybrid-VPIC simulations quench DCLC losses at higher $n_\mt{cool}/n_\mt{hot}$
and lower $T_\mt{cool}$ than predicted by the linear theory.
The hot-ion confinement is worse with $T_\mt{cool} = 5 \unit{keV}$ as compared
to lower $T_\mt{cool}$ (Figure~\ref{fig:cool-confinement}(f)-(h)),
which contrasts with Figure~\ref{fig:cool-plasma}'s prediction
that $T_\mt{cool} = 5 \unit{keV}$ with $n_\mt{cool} \approx n_\mt{hot}$
fully stabilizes DCLC within a wide $(k,\Real(\omega),\Imag(\omega))$ domain.
Why does $T_\mt{cool} \sim 1 \unit{keV}$ work better than $5 \unit{keV}$?
It may be explained by some combination of
(i) weaker electrostatic trapping and faster outflow $v_\mt{ti\prll}$ as
$T_\mt{cool}$ increases, and
(ii) quasi-linear diffusion of beam ions towards the loss cone, which shifts
the unstable drive $\dtl F/\dtl v_\perp$ to lower $v_\perp$ so that lower
$T_\mt{cool}$ becomes stabilizing.

Let us expand on point (ii).
For a quasi-linearly diffused $F(v_\perp)$, the relevant $T_\mt{cool}$ is set
not by the injected beam distribution, but instead by the loss-cone's $v_\perp$
boundary value \emph{at the injected beam's characteristic $v_\parallel$}.
For WHAM's $45^\circ$ pitch-angle beam, DCLC-scattered ions escape at the
loss-cone boundary with perpendicular energy
\begin{equation} \label{eq:wham-Tcool}
    \mi v_\perp^2 / 2
    \sim E_\mt{beam} \cos^2\theta_\mt{NBI}/(\mratio-1)
    \approx 0.7 \unit{keV}
    \, ,
\end{equation}
using $E_\mt{beam} = 25 \unit{keV}$, $\theta_\mt{NBI} = 45^\circ$,
and $\mratio=20$.
Equation~\ref{eq:wham-Tcool} agrees with the $T_\mt{cool} = 1 \unit{keV}$
stabilization in our simulations (Figure~\ref{fig:cool-confinement}).

It is interesting to contrast WHAM with 2XIIB, which used $90^\circ$
pitch-angle beam injection.
In 2XIIB, DCLC-scattered ions have $v_\prll \sim 0$ and therefore escape at the
loss-cone boundary with perpendicular energy
\begin{equation} \label{eq:2XIIB-Tcool}
    \mi v_\perp^2 / 2 \sim q_\mt{i} \Delta \phi /(\mratio - 1)
    \sim 0.2\text{--}0.5 \unit{keV}
\end{equation}
using $q_\mt{i} \Delta \phi \sim (2\text{--}5) \times \Te$,
$\Te \approx 100 \unit{eV}$, and $\mratio = 2$ \citep{coensgen1975};
here $\Delta\phi$ is the axial potential drop from mid-plane to throat
\citep[{\S}V.B]{baldwin1977}.
So, we may bear in mind that the appropriate $T_\mt{cool}$ to stabilize DCLC
(and the parametric dependence of $T_\mt{cool}$ upon either $E_\mt{beam}$ or
$\Delta \phi$) is mediated by the beam injection angle in a given device.

\subsection{Spatial Gradient Effects} \label{sec:gradient}

\begin{figure*}
    \centering
    \includegraphics[width=\textwidth]{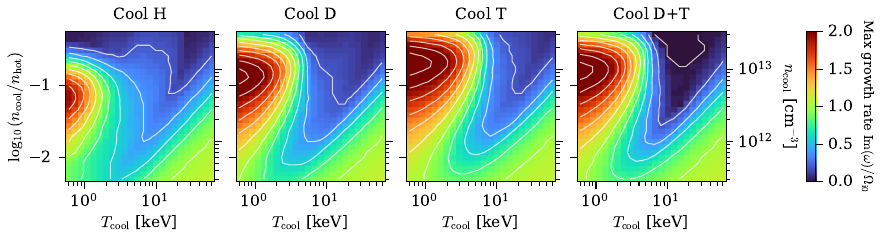}
    \caption{
        Effect of cool plasma on DCLC linear stability in a
        \emph{physically-larger} next-step mirror, similar to the BEAM concept
        described in \citet{forest2024}, with spatial gradient
        $|\epsilon| \rLio = 0.04$ smaller than in WHAM.
        Each panel shows varying cool plasma composition.
        For the cool D+T case, $n_\mt{cool}$ counts both D/T species, and the
        cool D and cool T have equal densities.
        Total stabilization $\Imag(\omega) \to 0$ is achieved when the cool ions'
        isotopes are matched to that of the hot ions.
        Colormap range in $\Imag(\omega)$ is reduced
        from Figure~\ref{fig:cool-plasma}(a).
    }
    \label{fig:cool-plasma-BEAM}
\end{figure*}

A smaller spatial gradient $\epsilon \rLio$ also helps to stabilize DCLC
\citep{baldwin1977,correll1980,ferron1984}.
Figure~\ref{fig:cool-plasma-BEAM} shows this for
plasma parameters similar to the physically-larger,
Break-Even Axisymmetric Mirror (\emph{BEAM}) design
concept of \citet{forest2024}.
We re-compute DCLC linear stability for \emph{BEAM}'s radial edge comprising
(i) hot deuterium/tritium (D/T) beam ions, with equal densities of D/T and
temperature $T \approx 60 \unit{keV}$ \citep[Fig.~6]{forest2024}, and
(ii) cool Maxwellian ions with varying $n_\mt{cool}$, $T_\mt{cool}$, and
isotope choice of hydrogen, deuterium, tritium, or a D/T mixture (equal
densities of D/T).
The stability calculation assumes $\epsilon \rLio = -0.04$, $B = 3 \unit{T}$,
and $n_\mt{hot} = 6 \times 10^{13} \unit{cm^{-3}}$ counting both D/T species.
The value $1/|\epsilon\rLio| = 25$ approximately matches the DCLC design
constraint $a/\rLi = 25$ used by both \citet{forest2024,frank2024}.
For normalization, we take $\rLio = 1.2 \unit{cm}$ and
$f_\mt{ci0} = 22.9 \unit{MHz}$.
We solve Equation~\eqref{eq:dclc-coldlec} using the same procedure as in
{\S}\ref{sec:dclc-id}, within a finite
domain $k\rLio < 30$, $\Real(\omega)/\Omcio < 10$, and $\Imag(\omega)/\Omcio < 5$.
The domain is larger than before because DCLC appears at larger $k \rLio$;
the relevant $k$ may be estimated for the $n$-th cyclotron harmonic as
$k \rLio \approx n/(|\epsilon| \rLio)/(T_\mt{i\perp}/T_\mt{i0})$, from
requiring that the ion diamagnetic drift $\omega/k = v_\mt{Di}$ intersects the
harmonics $\omega = n\Omcio$.
The Bessel sums retain all terms with index $|n| \le 120$ to ensure convergence.

Figure~\ref{fig:cool-plasma-BEAM} predicts that a larger region of the
parameter space $(n_\mt{cool}/n_\mt{hot}, T_\mt{cool}/T_\mt{hot})$ becomes
available to help stabilize DCLC in a BEAM-like concept.
Complete stabilization $\Imag(\omega) \to 0$ occurs when the cool plasma is a
D/T mixture like the hot plasma, following the empirical ``spectral rule'' of
\citet{kotelnikov2018}.
For cool plasma of pure hydrogen, deuterium, or tritium, we find that
$\Imag(\omega)$ is reduced but generally remains non-zero; the remaining
unstable modes have $\omega$ at the hot-ion cyclotron harmonics not overlapped
by the cool-ion harmonics, as previously shown by \citet{kotelnikov2018}.

Both \citet[Fig.~1]{tang1972} and \citet[Fig.~7]{baldwin1977} also computed the
maximum radial gradient $\epsilon$ for DCLC to be stable, as a function of
the density-proxy parameter $(\Omci/\ompi)^2$.
For WHAM,
$\epsilon\rLio \sim 1$ is DCLC unstable for nearly all values of $(\Omci/\ompi)^2$ anyways.
For the model BEAM plasma in Figure~\ref{fig:cool-plasma-BEAM}, we find
$(\Omci/\ompi)^2 \sim 3 \times 10^{-4}$ requires low $\epsilon\rLio \sim 0.01$
for stability, so it is reasonable that our model with $|\epsilon\rLio| \sim
0.04$ remains DCLC unstable in the absence of cool plasma.

Though Figure~\ref{fig:cool-plasma-BEAM} suggests that a BEAM-like concept may
be DCLC unstable, we note that many mitigating factors remain.
First, BEAM-sized plasmas need much lower $n_\mt{cool}$ to stabilize DCLC as
compared to WHAM, as expected from previous work \citep{baldwin1977};
there are many ways to craft such cool plasma in the laboratory.
Second, the peaked beam-ion distributions used here may be viewed as
``maximally'' unstable; quasi-linear diffusion will smooth ion distributions
towards marginal stability, as discussed in \S\ref{sec:confinement} and
\S\ref{sec:cool}.
Third, our calculation neglects physical effects such as finite plasma $\beta$
(i.e., $\del B$ along $r$) and both radial and axial geometry; these effects
are generally thought to aid stability \citep{tang1972}.
Fourth, recall from Figure~\ref{fig:cool-plasma} that even if DCLC remains
unstable, it can be rendered less harmful by pushing $\Real(\omega)$ to high
harmonics of $\Omcio$ and so reducing DCLC's scattering rate, as shown on the
GDT device \citep{shmigelsky2024-dclc}.
Fifth, the plasma parameters in Figure~\ref{fig:cool-plasma-BEAM} are only an
example; no attempt was made, for this manuscript, to optimize parameters
beyond what was discussed in \citet{forest2024}.
Lastly, we recall that DCLC has been successfully mitigated in past and current
mirror devices, including two that used WHAM/BEAM-like sloshing-ion injection:
TMX-U and GDT.

As an aside: the 2D parameter-regime maps of Figures~\ref{fig:cool-plasma}(a)
and \ref{fig:cool-plasma-BEAM} show interesting structure that has been studied
in detail by \citet[Figures~1,3,4]{gerver1976};
Gerver used a subtracted-Maxwellian distribution for hot ions, unlike our
arbitrary beam-ion distributions, but his results agree qualitatively with
ours.
For example, Figure~\ref{fig:cool-plasma-BEAM} shows that at low
$T_\mt{cool}/T_\mt{hot}$, a distinct instability occurs even at large
$n_\mt{cool}/n_\mt{hot} \gtrsim 10^{-1}$;
it is called double-humped instability by \citet{baldwin1977,kotelnikov2017}
or ion two-temperature instability by \citet{gerver1976}.
The interested reader may consult \citet{gerver1976,baldwin1977,post1987-review,kotelnikov2017}
for more thorough treatments and reviews of DCLC linear-stability physics.

\subsection{Kinetic Electron Effects} \label{sec:ELD}

Our linear dispersion relation assumed $k = k_\perp$, neglecting both ion and
electron parallel responses.
But, $k_\parallel \sim \pi/(2 L_\mt{p})$ is imposed by the mirror geometry for
the lowest possible axial harmonic.
In WHAM with $\mratio=20$, electrons with $T_\mt{e} \sim 1 \unit{keV}$ have
thermal velocity $v_\mt{te}$ similar to DCLC parallel phase velocity
$\omega/k_\parallel \sim \Omcio/k_\parallel$, so DCLC modes may be Landau damped by electrons.

We qualitatively assess the effect of parallel electron kinetics in
Equation~\eqref{eq:dclc-coldlec} by replacing the perpendicular, cold-fluid
electron susceptibility:
\begin{equation} \label{eq:chi-lec-to-replace}
    \chi_\mt{e} =
    \frac{\ompe^2}{\Omce^2}
    + \frac{\ompe^2}{|\Omce|}
    \frac{\epsilon}{k \omega}
    \, ,
\end{equation}
with a more general form for oblique electrostatic waves
that includes a $\vec{B}$-parallel kinetic response:
\begin{equation} \label{eq:chi-PKPM-maintext}
    \chi_\mt{e} =
        -
        \left(\frac{k_\perp}{k}\right)^2
        \frac{\ompe^2}{\Omce^2}
        \left[
            1
            + \frac{\epsilon|\Omce|}{k_\perp \omega}
        \right]
        \zeta_{0\mt{e}} Z(\zeta_{0\mt{e}})
        - \left(\frac{k_\parallel}{k}\right)^2
        \frac{\ompe^2}{k_\prll^2 v_\mt{te}^2}
        Z'(\zeta_{0\mt{e}})
    \, .
\end{equation}
Here $Z$ is the plasma dispersion function,
$\zeta_{0\mt{e}} = \omega/(k_\prll v_\mt{te})$,
and $v_\mt{te} = \sqrt{2 T_\mt{e}/m_\mt{e}}$.
We fix $k_\parallel = \pi/(2 L_\mt{p})$ to mimic a fundamental-harmonic
mode along the device axis.
Both Equations~\eqref{eq:chi-lec-to-replace} and \eqref{eq:chi-PKPM-maintext}
are dimensionful.
The derivation is briefly sketched in Appendix~\ref{app:electrons}.

\begin{figure}
    \centering
    \includegraphics[width=0.85\textwidth]{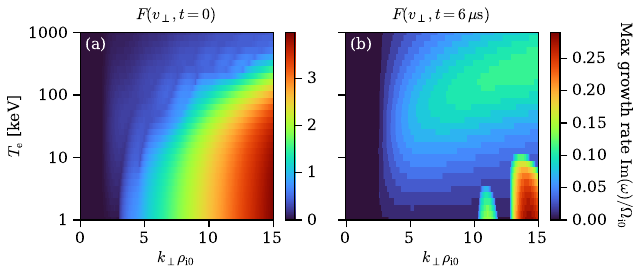}
    \caption{
        Effect of parallel-kinetic electron response upon DCLC linear
        stability, using ion distributions from the WHAM $\mratio=20$
        simulation at either $t=0$ to obtain a beam-ion distribution (left),
        or at $t=6\unit{\mu s}$ to obtain a saturated distribution with
        $\dtl F/\dtl v_\perp < 0$ (right).
    }
    \label{fig:landau-damping}
\end{figure}

Figure~\ref{fig:landau-damping} re-computes DCLC linear stability, using
Equation~\eqref{eq:chi-PKPM-maintext} to show the effect of parallel electron
kinetics, for the ion distributions from our WHAM $\mratio=20$ simulation at
$t=0$ and $t=6\,\tbounce \approx 6\unit{\mu s}$.
Figure~\ref{fig:landau-damping}(a) shows that the $t=0$, peaked beam-ion
distribution with empty loss cone remains unstable for a broad range of $k$;
electron kinetics do not stabilize a strongly-peaked and hence
strongly-unstable $F(v_\perp)$.
In contrast, Figure~\ref{fig:landau-damping}(b) shows that the
marginally-unstable $t=6\unit{\mu s}$ distribution with filled loss cone has
DCLC growth rates reduced by electron kinetics.

In the limit $\Te \to \infty$, $\zeta_{0\mt{e}} \to 0$ suppresses the electron
parallel susceptibility;
i.e., the $(k_\parallel/k)^2$ term in Equation~\eqref{eq:chi-PKPM-maintext}
asymptotes to $1/(k^2 \lde^2)$, where $\lde$ is the electron Debye length,
and its magnitude and contribution to $D$ is negligible.
More importantly, hot electrons and finite $k_\parallel$ suppress the
\emph{perpendicular} drift term in Equation~\eqref{eq:chi-PKPM-maintext} by
driving $\zeta_{0\mt{e}} Z(\zeta_{0\mt{e}}) \to 0$;
this disables the coupling between ion Bernstein waves and the drift wave.
In Figure~\ref{fig:landau-damping}(b), at $\Te \gtrsim 50 \unit{keV}$ the
resulting mode structure appears similar to the ``pure'' ion Bernstein waves in
a homogeneous plasma, but with non-zero growth rates.
These unstable Bernstein modes are negative-energy waves satisfying the
criterion
$\ptl\left[\omega \Real(D)\right]/\ptl \omega|_{\Real(\omega)} <0$
\citep[\S4-2]{stix1992};
in this criterion we replaced the Hermitian part of the dielectric tensor with
$\Real(D) = \Real(1 + \chi_\mt{i} + \chi_\mt{e})$,
which is valid because $D$ has no contributions from off-diagonal terms
$\chi_{\parallel\perp}$ in either species' susceptibility tensor.
See also \citet{kadomtsev1964,bers1965,baldwin1977}.

In the limit of low $\Te \lesssim 1 \unit{keV}$, the perpendicular drift term
in Equation~\eqref{eq:chi-PKPM-maintext} reverts to its fluid form because
$\zeta_{0\mt{e}} Z(\zeta_{0\mt{e}}) \to -1$.
The parallel term, which asymptotes to $-(k_\prll/k)^2(\ompe/\omega)^2$,
is the main new influence on DCLC mode structure.
Figure~\ref{fig:landau-damping}(a) shows that the $t=0$ beam-ion distribution
is not much affected at low $\Te$ when compared to Figure~\ref{fig:mode-f}(j).
But, Figure~\ref{fig:landau-damping}(b) shows that the $t=6\unit{\mu s}$
distribution has low harmonics of DCLC suppressed, and the growth rates of
higher harmonics somewhat reduced, by electron kinetics when compared to
Figure~\ref{fig:mode-f}(g).

Equation~\eqref{eq:chi-PKPM-maintext} is less accurate than bounce averaging of unperturbed particle
orbits within a specified axial mirror geometry, as has been performed and
studied by, e.g., \citet{cohen1983,koepke1986-BRLD}, and others.
A significant unknown is the effect of the non-monotonic axial electric
potential $\phi$; since $\phi \sim \Te$ and $\phi$ can trap electrons at
sloshing-ion turning points, electron orbits may be significantly modified.
None of this is captured in our Hybrid-VPIC simulations given the simple
electron closure.
Our goal is only to show qualitatively how parallel electron kinetics,
including electron Landau damping, may impact DCLC.
We conclude that saturated DCLC amplitude and frequency in WHAM
may be tunable via $\Te$ or other device parameters, as was done on the MIX-1
device previously \citep{koepke1986-BRLD,koepke1992}.

\subsection{Other Modes} \label{sec:other-modes}

Our simulations mostly grow DCLC, but other kinetic and fluid modes can appear
in mirror devices \citep{post1987-review}.
The modes relevant to WHAM were surveyed by \citet{endrizzi2023};
here we add a few remarks.

\begin{figure*}
    \centering
    \includegraphics[width=\textwidth]{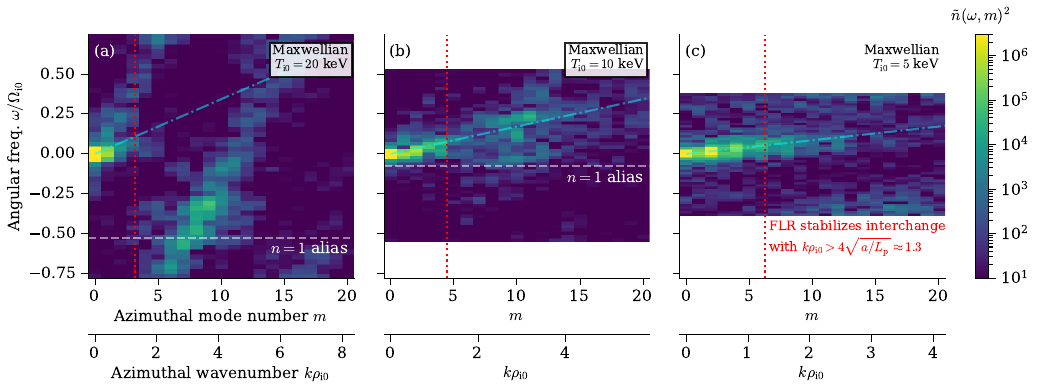}
    \caption{
        Interchange modes appear and DCLC weakens as $\Ti$ decreases (left to
        right) in simulations of Maxwellian ions in WHAM's $\mratio=20$
        magnetic-field geometry.
        Fourier spectra computed as in Figure~\ref{fig:mode-f}.
        The dot-dashed cyan line plots the interchange mode's expected phase
        velocity, $\omega/k = v_\mt{Di}/2$, assuming spatial gradient
        $\epsilon = (10 \unit{cm})^{-1}$.
    }
    \label{fig:interchange}
\end{figure*}

Interchange modes should be stabilized by the effect of finite ion Larmor
radius (FLR), specifically collisionless gyroviscosity \citep{roberts1962},
when
\begin{equation} \label{eq:interchange-flr}
    k_\perp \rLio > 4 \sqrt{ \frac{a}{L_\mt{p}} } \approx 1.3
    \, ,
\end{equation}
with $a \approx 10 \unit{cm}$ the plasma column radius and assuming a
curvature-driven growth rate
$\vtio / \sqrt{a L_\mt{p}}$ \citep[{\S}III.B]{ryutov2011}.
In Figure~\ref{fig:interchange}, we present new simulations of Maxwellian ions
with varying temperature $\Tio = 5$ to $20 \unit{keV}$ in the WHAM $\mratio=20$
geometry.
As $\Tio$ decreases, DCLC weakens in amplitude and spectral width,
and a lower-$m$ mode grows in amplitude and spectral width.
We identify the lower-$m$ mode as interchange because
(i) its phase velocity is half the ion diamagnetic drift and in the same
direction (neglecting gravitational drift, which is $\abt (a/L_\mt{p}) \times$
smaller), consistent with the planar-slab derivation
\citep{rosenbluth1962,roberts1962},
(ii) its $k$ bandwidth qualitatively scales with $\Ti$ following
Equation~\eqref{eq:interchange-flr}.
Note that in this paragraph and Figure~\ref{fig:interchange},
we redefine $\vtio = \sqrt{2\Tio/\mi}$ with a factor of $\sqrt{2}$, which
also affects $\rLio = \vtio/\Omcio$.
We caution that our simulated interchange has relatively strong $m=2$ and $m=4$
modes compared to, e.g., the odd $m=3$ mode; this effect may be unphysical, and
we suspect mesh imprinting.

Alfv\'{e}n ion cyclotron (AIC) modes do not appear at significant amplitude in
our simulations; recall that both $\delta B_r$ and $\delta B_\theta$ are small
({\S}\ref{sec:results-overview}), and pitch-angle scattering is weak compared to
DCLC's $v_\perp$ scattering (Figure~\ref{fig:vdfs}).
Does our non-observation agree with theory and prior experiments?
An empirical criterion for AIC growth, obtained from experiments on the tandem
mirror GAMMA-10 \citep{ichimura1993,katsumata1996}, is
\begin{equation} \label{eq:aic-gamma10}
    T_{\mt{i}\perp}/T_{\mt{i}\parallel} > 0.55/\beta_{\perp}^{0.5}
    \, ,
\end{equation}
based on data with $\beta_\perp < 0.03$.
One linear-instability criterion, derived for a homogeneous bi-Maxwellian plasma
\citep{gary1994-ic-sim} and with a form commonly used in solar-wind literature
\citep[e.g.,][]{hellinger2006}, is:
\begin{equation} \label{eq:aic-gary}
    T_{\mt{i}\perp}/T_{\mt{i}\parallel} > 1 + 0.43/\beta_{\prll}^{0.42}
    \, .
\end{equation}
At $t=0$ in our $\mratio=20$ simulation, the sloshing-ion turning points have
$\beta_{\perp} = 8\pi n T_{\mt{i}\perp} / B^2 = 0.17$
and $\beta_{\prll} = 8\pi n T_{\mt{i}\prll} / B^2 = 0.068$,
with corresponding anisotropy $T_{\mt{i}\perp}/T_{\mt{i}\prll} = 2.5$;
our simulations with larger $\mratio$ have similar anisotropy and lower plasma
beta at the turning points.
Both Equations~\eqref{eq:aic-gamma10} and \eqref{eq:aic-gary} indicate that AIC
may be unstable at the turning points.

So, why does AIC not appear?
First, since AIC is driven by gradients of $f$ on resonant surfaces in velocity
space, Equations~\eqref{eq:aic-gamma10} and \eqref{eq:aic-gary} will not be so
precise when applied to different ion distributions; e.g.,
\citet{isenberg2013} noted that subtle modifications to $f$ at marginal AIC
stability can modify anisotropy thresholds based on bi-Maxwellian temperatures
by a factor of $\abt 2$.
Second, AIC is stabilized by the inhomogeneous plasma in WHAM.
Sloshing ions put perpendicular pressure anisotropy at turning points, so
instability drive weakens towards the mirror cell's center.
A small plasma column radius with respect to the ion Larmor radius also aids
stability \citep{tsidulko2014}.
And, AIC is suppressed if the mirror's axial length is shorter than a critical
length \citep{tajima1977,tajima1980,nicks2023}:
\begin{equation} \label{eq:aic-length}
    L_\mt{c} =
    2 \pi^2 \sqrt{ \frac{T_{\mt{i}\prll}}{\beta_\perp T_{\mt{i}\perp}} }
    \left( \frac{c}{\ompi} \right)
    \, .
\end{equation}
The critical length $L_\mt{c} \approx 182 \unit{cm}$ is very close to WHAM's length
$2 L_\mt{p} = 196 \unit{cm}$, again taking $\beta_\perp = 0.17$ and assuming
$c/\ompi=5.9\unit{cm}$ for density $n = 3 \times 10^{13} \unit{cm^{-3}}$ at
sloshing-ion turning points.
Third, DCLC simply has a faster growth rate and decreases plasma beta before
AIC can be triggered.

To summarize: AIC with low axial mode number may be marginally unstable for
WHAM, based on the highest possible $\beta_{\perp}$ and density $n$ at
sloshing-ion turning points.
But multiple effects weaken AIC drive and so may explain why it does not appear
in our simulations.

\subsection{Comparison to Real Mirrors} \label{sec:comparison}

In real mirror devices, discrete DCLC modes can persist stably for
$\abt \mathcal{O}(1 \unit{ms})$ \citep{shmigelsky2024-dclc}, but DCLC can also
appear as discrete bursts of enhanced fluctuations with duration $\abt 10$ to
$100 \unit{\mu s}$ \citep{coensgen1975,yamaguchi1996,shmigelsky2024-dclc}.
Our simulations do not show bursting, nor did previous simulations by
\citet{cohen1983}.
\citet{yamaguchi1996} explained bursty DCLC in the GAMMA-6A
experiment using a quasi-linear model with bounce-averaged electron Landau
damping; they appealed to
(i) separation between DCLC scattering and axial outflow timescales (i.e.,
$1/\nu_{\perp\perp} \ll \tau_\mt{GD}$), and
(ii) fast time variation in DCLC growth rate with slower variation in
electron-Landau damping rate.
In our simulations, $1/\nu_{\perp\perp} \sim \tau_\mt{GD}$ at order of
magnitude (Figure~\ref{fig:confinement-time}(b)); DCLC appears marginally
stable and does not damp upon ions within a timescale $\ll \tau_\mt{GD}$.
It would be interesting to see if future kinetic simulations with longer
simulation durations, electron Landau damping, or other physical effects can
replicate DCLC bursting.

TMX-U and GDT saw that DCLC could be driven at sloshing-ion turning
points instead of at the mirror mid-plane $z=0$
\citep{berzins1987,shmigelsky2024-dclc};
Why does DCLC have strongest drive at $z=0$, versus at the turning points, in
our simulations of WHAM?
In TMX-U, the end-plug could be stabilized on one side and not the other due to a combination of axial flows from the
central cell and localized ECH at the end-plug outer-turning point \citep{berzins1987}.
As for GDT versus WHAM, we cannot answer definitively, but we note that
WHAM's shorter axial length of $2\unit{m}$
may constrain DCLC's
axial eigenmodes as compared to
GDT's $7 \unit{m}$ length.

Where do cool ions come from in real experiments?
In WHAM, DCLC-scattered ions escape with large $|v_\parallel|$ and are not trapped.
Charge exchange between beam ions and cool neutrals can generate $< 1 \unit{keV}$
ions that trap in the mid-plane's electrostatic potential well;
prerequisite cool neutrals naturally outgas from plasma-facing materials.
In tandem mirrors like TMX-U, central-cell outflow into end-plug cells can also
provide cool ions to stabilize DCLC \citep{simonen1983,berzins1987}.
Central-cell outflow may also help to reduce the growth rate of
curvature-driven collisionless trapped-particle modes
\citep{rosenbluth1982,berk1983}, as outflowing ions can sample good curvature
near mirror throats and help to couple adjacent MHD-unstable and stable mirror
cells.

Cool-ion stabilization of DCLC may face problems in larger and more powerful mirrors.
The axial electrostatic well at $z=0$, which traps cool ions, will become
shallower due to ion-ion pitch-angle scattering of the beam ions.
Pitch-angle scattering is counter-balanced by neutral-beam capture via
charge exchange (CX) or impact ionization, as both processes create new hot
ions with $45^\circ$ pitch-angle.
CX between hot ions / fast neutrals also pumps out the isotropized,
scattered hot ions, so one CX event collimates the ions' pitch-angle
distribution more effectively than one impact-ionization event.
But, the CX cross section drops relative to impact ionization for beam
energies $\gtrsim 70 \unit{keV}$ \citep[Figure~10]{endrizzi2023}.
It thus becomes harder to collimate the ion pitch-angle distribution and
harder to maintain an axial electrostatic potential well with
$\mathcal{O}(10^2) \unit{keV}$ NBI
\citep[e.g.,][Figure~6]{forest2024}.
And, tandem central-cell outflow reduces fusion performance and may set limits on power-plant reactor design \citep[e.g.,][\S3.2]{frank2024}.
Future mirror designs may thus need
other methods to help stabilize DCLC.

WHAM began operating in July 2024 \citep{anderson2024}.
A comparison between experimental data and our simulations is not yet available.
We anticipate that WHAM plasmas and diagnostics may be tuned to create and
measure DCLC modes in future experimental campaigns.

\section{Conclusions}
\label{sec:conclusions}

We have performed 3D kinetic-ion simulations of WHAM, initialized with a
neutral-beam-injected deuteron population with $\Ti \sim 10 \unit{keV}$ and
cool, isothermal-fluid electrons with $\Te \sim 1 \unit{keV}$, to assess
kinetic plasma stability in a high-performance, collisionless-ion regime.
We find that WHAM's beam-ion distribution is unstable to an electrostatic,
flute-like ($k \approx k_\perp$) mode that grows on $\lesssim 1 \unit{\mu s}$
timescales; it propagates azimuthally around the column in the ion diamagnetic
direction and has angular frequency between $\Omcio$ and $2\Omcio$.
We identify it as the drift-cyclotron loss cone (DCLC) mode, well known from
prior mirror experiments \citep{coensgen1975} and previously anticipated to be
a possible concern for WHAM \citep{endrizzi2023}.

The plasma column and DCLC fluctuations settle into a steady-state decay by
$t = 6 \unit{\mu s}$.
Particles escape axially with confinement time
$\tloss = n / (\dtl n/\dtl t) \sim 10^2 \unit{\mu s}$ in a ``gas dynamic''
regime, wherein the scattering rate into the loss cone equals or exceeds the
rate of free-streaming axial loss from the mirror.
Particle losses are due to collisionless $v_\perp$ scattering upon the DCLC
modes; the particle-wave correlation time is approximately ${\Omcio}^{-1}$ at
the mirror mid-plane.
Particle losses and velocity-space diffusion are strongest at the plasma's
radial edge, whereas the plasma column's core can maintain $\dtl F/\dtl v_\perp > 0$ at low
$v_\perp$.

We review well-known and experimentally-tested methods for stabilizing DCLC:
addition of cool plasma to fill the loss cone,
larger plasma extent (smaller gradient),
and parallel electron kinetics including Landau damping.
In 3D simulations with cool ions initialized at the plasma's radial edge, the
beam ions' confinement time can be raised by up to $9\times$,
though an order-unity ratio of cool/hot ion number density is needed
and the cool-ion confinement is poorer than that of the beam ions.
The best-case beam-ion confinement time of several $100 \unit{\mu s}$ also
remains two orders of magnitude below the ideal, ``classical'' confinement
time of 0.1--0.2 seconds.
In a real experiment, the cool ions must be provided and replenished by
external sources because DCLC does not scatter beam ions into the axial
electrostatic potential well \citep{kesner1973,kesner1980}
where they could be trapped to help stabilize DCLC.
DCLC-scattered beam ions are lost because they retain large parallel speeds and
so never enter the trapped region of phase space.

Our simulations are limited, especially in the electric field model and
isothermal-fluid electron closure.
Future work may incorporate electron inertia, an electron energy equation,
drift-kinetic electrons, or more to help model (i) bounce-averaged electron
Landau damping, and (ii) the plasmas' axial and radial potential structure,
which dictate outflows and rotation.
And, the initial condition of a hot, beam-ion plasma with only mild
slowing-down upon electrons is idealized.
A two-way coupling between Hybrid-VPIC and CQL3D-m over a $20\unit{ms}$
laboratory shot duration, passing DCLC quasi-linear diffusion coefficients from
kinetic simulations into the Fokker-Planck equation, may yield more realistic
predictions.
Adding neutral-ion interactions to our CQL3D-m models would improve our
cool-ion stabilization modeling.
Other subsystems on WHAM are not modeled, e.g., heating of ions and
electrons via RF and microwave radiation respectively, or biased end-rings
within the expanders used to drive rotation and shear flow.
So, future work may also consider a wider range of fueling and heating
scenarios in WHAM and next-step mirror devices.

\section*{Acknowledgements}

Conversations with Dominick Bindl, Sean Dettrick, Mana Francisquez, Roelof Groenewald, Mark Koepke, Dmitri
Ryutov, Derek Sutherland, Dmitry Yakovlev, Mason Yu, and the entire WHAM team are gratefully acknowledged.
We also thank the anonymous referees for constructive comments that have improved this paper.

\section*{Funding}

Funding
for this work was
provided by Realta Fusion, the U.S. Department of Energy (DOE), and the
U.S. National Science Foundation (NSF).
WHAM collaboration work is supported by the U.S. DOE through ARPA-E DE-AR0001258,
Commonwealth Fusion Systems, Realta Fusion, Wisconsin Alumni Research
Foundation, and a lengthy list of collaborators providing valuable equipment.
AT was partly supported by NSF PHY-2010189 and the DOE Fusion Energy Sciences
Postdoctoral Research Program, administered by the Oak Ridge Institute for
Science and Education (ORISE) and Oak Ridge Associated Universities (ORAU)
under DOE contract DE-SC0014664.

High-performance computing resources were provided by
the National Energy Research Scientific Computing Center (NERSC), a DOE Office of Science User Facility, via allocation FES-ERCAP0026655;
Anvil at Purdue University's Rosen Center for Advanced
Computing \citep{mccartney2014,song2022} via allocation PHY230179 from the NSF Advanced
Cyberinfrastructure Coordination Ecosystem: Services \& Support (ACCESS)
program;
Amazon Web Services' Compute for Climate Fellowship awarded to Realta Fusion;
Los Alamos Institutional Computing;
and UW--Madison's Center for High Throughput Computing \citep{chtc2006}.
ACCESS \citep{boerner2023} is supported by NSF grants \#2138259, \#2138286,
\#2138307, \#2137603, and \#2138296.

All opinions expressed in this paper are the authors' and do not necessarily
reflect the policies and views of DOE, ORAU, or ORISE.

\section*{Declaration of interests}

C.~B.~Forest is a co-founder of Realta Fusion;
D.~A.~Endrizzi, S.~J.~Frank, and J.~Viola are employees of Realta Fusion.
This work was partly supported by a grant from Realta Fusion to the University
of Wisconsin--Madison.

\bibliographystyle{jpp}
\bibliography{library}

\begin{thebibliography}{90}
\expandafter\ifx\csname natexlab\endcsname\relax\def\natexlab#1{#1}\fi
\def\au#1{#1} \def\ed#1{#1} \def\yr#1{#1}\def\at#1{#1}\def\jt#1{\textit{#1}}
  \def\bt#1{#1}\def\bvol#1{\textbf{#1}} \def\vol#1{#1} \def\pg#1{#1}
  \def\publ#1{#1}\def\arxiv#1{#1}\def\org#1{#1}\def\st#1{\textit{#1}}

\bibitem[{Aamodt}(1977)]{aamodt1977-electron}
{\sc \au{{Aamodt}, R.~E.}} \yr{1977}  \at{{Electron stabilization of drift-cone
  modes}}.  \jt{Physics of Fluids}  \bvol{20}~(6),  \pg{960--962}.

\bibitem[{Aamodt} {\em et~al.\/}(1981){Aamodt}, {Cohen}, {Lee}, {Liu},
  {Nicholson} \& {Rosenbluth}]{aamodt1981}
{\sc \au{{Aamodt}, R.~E.}, \au{{Cohen}, B.~I.}, \au{{Lee}, Y.~C.}, \au{{Liu},
  C.~S.}, \au{{Nicholson}, D.~R.} \& \au{{Rosenbluth}, M.~N.}} \yr{1981}
  \at{{Nonlinear evolution of drift cyclotron modes}}.  \jt{Physics of Fluids}
  \bvol{24}~(1),  \pg{55--65}.

\bibitem[{Aamodt} {\em et~al.\/}(1977){Aamodt}, {Lee}, {Liu} \&
  {Rosenbluth}]{aamodt1977-shift}
{\sc \au{{Aamodt}, R.~E.}, \au{{Lee}, Y.~C.}, \au{{Liu}, C.~S.} \&
  \au{{Rosenbluth}, M.~N.}} \yr{1977}  \at{{Nonlinear dynamics of
  drift-cyclotron instability}}.  \jt{\prl}  \bvol{39}~(26),  \pg{1660--1663}.

\bibitem[{Anderson} {\em et~al.\/}(2024){Anderson}, {Anderson}, {Biewer},
  {Biswas}, {Brown}, {Claveau}, {Clark}, {Egedal}, {Endrizzi}, {Forest},
  {Fujii}, {Furlong}, {Frank}, {Ialovega}, {Kirch}, {Kristofek}, {Gonzalez},
  {Lindley}, {Marriott}, {Mirnov}, {Mumgaard}, {Murdock}, {Penne}, {Pizzo},
  {Oliva}, {Qian}, {Sanwalka}, {Schmitz}, {Shih}, {Tran}, {Viola}, {Wallace},
  {Yakovlev} \& {Yu}]{anderson2024}
{\sc \au{{Anderson}, J.~K.}, \au{{Anderson}, O.}, \au{{Biewer}, T.~M.},
  \au{{Biswas}, B.}, \au{{Brown}, M.~R.}, \au{{Claveau}, E.}, \au{{Clark}, M.},
  \au{{Egedal}, J.}, \au{{Endrizzi}, D.}, \au{{Forest}, C.~B.}, \au{{Fujii},
  K.}, \au{{Furlong}, K.}, \au{{Frank}, S.~J.}, \au{{Ialovega}, M.},
  \au{{Kirch}, J.}, \au{{Kristofek}, G.}, \au{{Gonzalez}, X.~N.},
  \au{{Lindley}, B.}, \au{{Marriott}, E.}, \au{{Mirnov}, V.}, \au{{Mumgaard},
  R.~T.}, \au{{Murdock}, S.}, \au{{Penne}, E.}, \au{{Pizzo}, J.~D.},
  \au{{Oliva}, S.~F.}, \au{{Qian}, T.}, \au{{Sanwalka}, K.}, \au{{Schmitz},
  O.}, \au{{Shih}, K.}, \au{{Tran}, A.}, \au{{Viola}, J.~D.}, \au{{Wallace},
  J.~P.}, \au{{Yakovlev}, D.} \& \au{{Yu}, M.}} \yr{2024} First physics results
  from the wisconsin hts axisymmetric mirror (wham).  \bt{In {\em APS Division
  of Plasma Physics Meeting Abstracts\/}},  \st{APS Meeting Abstracts},
  \vol{vol. 2024},  \pg{p. ZI03.00001}.

\bibitem[{Bagryansky}(2024)]{bagryansky2024}
{\sc \au{{Bagryansky}, P.~A.}} \yr{2024}  \at{{Progress of open systems at
  Budker Institute of Nuclear Physics}}.  \jt{Journal of Plasma Physics}
  \bvol{90}~(2),  \pg{905900218}.

\bibitem[{Bajborodov} {\em et~al.\/}(1971){Bajborodov}, {Ioffe}, {Kanaev},
  {Sobolev} \& {Jushmanov}]{bajborodov1971}
{\sc \au{{Bajborodov}, J.~T.}, \au{{Ioffe}, M.~S.}, \au{{Kanaev}, B.~I.},
  \au{{Sobolev}, R.~I.} \& \au{{Jushmanov}, E.~E.}} \yr{1971} {Investigation of
  Plasma Decay in the PR-6 Adiabatic Trap}.  \bt{In {\em Proceedings of the
  Fourth International Conference on Plasma Physics and Controlled Nuclear
  Fusion Research\/}}, ,  \vol{vol.~2}.  \publ{Vienna: International Atomic
  Energy Agency (IAEA)}.

\bibitem[{Baldwin}(1977)]{baldwin1977}
{\sc \au{{Baldwin}, D.~E.}} \yr{1977}  \at{{End-loss processes from mirror
  machines}}.  \jt{Reviews of Modern Physics}  \bvol{49}~(2),  \pg{317--339}.

\bibitem[{Baldwin} {\em et~al.\/}(1976){Baldwin}, {Berk} \&
  {Pearlstein}]{baldwin1976}
{\sc \au{{Baldwin}, D.~E.}, \au{{Berk}, H.~L.} \& \au{{Pearlstein}, L.~D.}}
  \yr{1976}  \at{{Turbulent Lifetimes in Mirror Machines}}.  \jt{\prl}
  \bvol{36}~(17),  \pg{1051--1054}.

\bibitem[{Beklemishev} {\em et~al.\/}(2010){Beklemishev}, {Bagryansky},
  {Chaschin} \& {Soldatkina}]{beklemishev2010}
{\sc \au{{Beklemishev}, A.~D.}, \au{{Bagryansky}, P.~A.}, \au{{Chaschin},
  M.~S.} \& \au{{Soldatkina}, E.~I.}} \yr{2010}  \at{{Vortex Confinement of
  Plasmas in Symmetric Mirror Traps}}.  \jt{Fusion Science and Technology}
  \bvol{57}~(4),  \pg{351--360}.

\bibitem[{Berk} {\em et~al.\/}(1983){Berk}, {Rosenbluth}, {Wong}, {Antonsen} \&
  {Baldwin}]{berk1983}
{\sc \au{{Berk}, H.~L.}, \au{{Rosenbluth}, M.~N.}, \au{{Wong}, H.~V.},
  \au{{Antonsen}, T.~M.} \& \au{{Baldwin}, D.~E.}} \yr{1983}  \at{{Fast growing
  trapped-particle modes in tandem mirrors}}.  \jt{Soviet Journal of Plasma
  Physics}  \bvol{9}~(1),  \pg{108}.

\bibitem[{Berk} \& {Stewart}(1977)]{berk1977}
{\sc \au{{Berk}, H.~L.} \& \au{{Stewart}, J.~J.}} \yr{1977}  \at{{Quasi-linear
  transport model for mirror machines}}.  \jt{Physics of Fluids}
  \bvol{20}~(7),  \pg{1080--1088}.

\bibitem[{Bers} \& {Gruber}(1965)]{bers1965}
{\sc \au{{Bers}, A.} \& \au{{Gruber}, S.}} \yr{1965}  \at{{Negative-Energy
  Plasma Waves and Instabilities at Cyclotron Harmonics}}.  \jt{Applied Physics
  Letters}  \bvol{6}~(2),  \pg{27--28}.

\bibitem[{Berzins} \& {Casper}(1987)]{berzins1987}
{\sc \au{{Berzins}, L.~V.} \& \au{{Casper}, T.~A.}} \yr{1987}  \at{{Ion
  microinstability at the outer sloshing-ion turning point of the tandem mirror
  experiment upgrade (TMX-U)}}.  \jt{\prl}  \bvol{59}~(13),  \pg{1428--1431}.

\bibitem[Boerner {\em et~al.\/}(2023)Boerner, Deems, Furlani, Knuth \&
  Towns]{boerner2023}
{\sc \au{Boerner, T.~J.}, \au{Deems, S.}, \au{Furlani, T.~R.}, \au{Knuth,
  S.~L.} \& \au{Towns, J.}} \yr{2023} Access: Advancing innovation: Nsf's
  advanced cyberinfrastructure coordination ecosystem: Services \& support.
  \bt{In {\em Practice and Experience in Advanced Research Computing\/}}, {\em
  PEARC '23\/} ,  \pg{pp. 173--176}.  \publ{New York, NY, USA: Association for
  Computing Machinery}.

\bibitem[{Bowers} {\em et~al.\/}(2008){Bowers}, {Albright}, {Yin}, {Bergen} \&
  {Kwan}]{bowers2008}
{\sc \au{{Bowers}, K.~J.}, \au{{Albright}, B.~J.}, \au{{Yin}, L.},
  \au{{Bergen}, B.} \& \au{{Kwan}, T.~J.~T.}} \yr{2008}  \at{{Ultrahigh
  performance three-dimensional electromagnetic relativistic kinetic plasma
  simulationa)}}.  \jt{Physics of Plasmas}  \bvol{15}~(5),  \pg{055703}.

\bibitem[{Burkhart} {\em et~al.\/}(1989){Burkhart}, {Guzdar} \&
  {Koepke}]{burkhart1989}
{\sc \au{{Burkhart}, G.~R.}, \au{{Guzdar}, P.~N.} \& \au{{Koepke}, M.~E.}}
  \yr{1989}  \at{{Theoretical modeling of drift cyclotron loss-cone instability
  mode structures}}.  \jt{Physics of Fluids B}  \bvol{1}~(3),  \pg{570--580}.

\bibitem[{Center for High Throughput Computing}(2006)]{chtc2006}
{\sc \au{{Center for High Throughput Computing}}} \yr{2006} Center for high
  throughput computing.

\bibitem[{Chaudhry}(1983)]{chaudhry1983}
{\sc \au{{Chaudhry}, M.~B.}} \yr{1983}  \at{{Electrostatic Drift Ion Cyclotron
  Waves in Sheet Plasmas with and without Ambipolar Field}}.  \jt{Journal of
  the Physical Society of Japan}  \bvol{52}~(3),  \pg{856}.

\bibitem[{Coensgen} {\em et~al.\/}(1975){Coensgen}, {Cummins}, {Logan},
  {Molvik}, {Nexsen}, {Simonen}, {Stallard} \& {Turner}]{coensgen1975}
{\sc \au{{Coensgen}, F.~H.}, \au{{Cummins}, W.~F.}, \au{{Logan}, B.~G.},
  \au{{Molvik}, A.~W.}, \au{{Nexsen}, W.~E.}, \au{{Simonen}, T.~C.},
  \au{{Stallard}, B.~W.} \& \au{{Turner}, W.~C.}} \yr{1975}  \at{{Stabilization
  of a Neutral-Beam-Sustained, Mirror-Confined Plasma}}.  \jt{\prl}
  \bvol{35}~(22),  \pg{1501--1503}.

\bibitem[{Cohen} \& {Maron}(1980)]{cohen1980}
{\sc \au{{Cohen}, B.~I.} \& \au{{Maron}, N.}} \yr{1980}  \at{{Simulation of
  drift-cone modes}}.  \jt{Physics of Fluids}  \bvol{23}~(5),  \pg{974--980}.

\bibitem[{Cohen} {\em et~al.\/}(1984){Cohen}, {Maron} \& {Nevins}]{cohen1984}
{\sc \au{{Cohen}, B.~I.}, \au{{Maron}, N.} \& \au{{Nevins}, W.~M.}} \yr{1984}
  \at{{Simulation of drift-cyclotron-loss-cone modes in tandem mirrors with
  sloshing ions}}.  \jt{Physics of Fluids}  \bvol{27}~(3),  \pg{642--649}.

\bibitem[{Cohen} {\em et~al.\/}(1982){Cohen}, {Maron} \& {Smith}]{cohen1982}
{\sc \au{{Cohen}, B.~I.}, \au{{Maron}, N.} \& \au{{Smith}, G.~R.}} \yr{1982}
  \at{{Some nonlinear properties of drift-cyclotron modes}}.  \jt{Physics of
  Fluids}  \bvol{25}~(5),  \pg{821--841}.

\bibitem[{Cohen} {\em et~al.\/}(1983){Cohen}, {Smith}, {Maron} \&
  {Nevins}]{cohen1983}
{\sc \au{{Cohen}, B.~I.}, \au{{Smith}, G.~R.}, \au{{Maron}, N.} \&
  \au{{Nevins}, W.~M.}} \yr{1983}  \at{{Particle simulations of ion-cyclotron
  turbulence in a mirror plasma}}.  \jt{Physics of Fluids}  \bvol{26}~(7),
  \pg{1851--1865}.

\bibitem[{Correll} {\em et~al.\/}(1980){Correll}, {Clauser}, {Coensgen},
  {Cummins}, {Drake}, {Foote}, {Futch}, {Goodman}, {Grubb} \&
  {Melin}]{correll1980}
{\sc \au{{Correll}, D.~L.}, \au{{Clauser}, J.~H.}, \au{{Coensgen}, F.~H.},
  \au{{Cummins}, W.~F.}, \au{{Drake}, R.~P.}, \au{{Foote}, J.~H.}, \au{{Futch},
  A.~H.}, \au{{Goodman}, R.~K.}, \au{{Grubb}, D.~P.} \& \au{{Melin}, G.~M.}}
  \yr{1980}  \at{{Production of large-radius, high-beta, confined mirror
  plasmas}}.  \jt{Nuclear Fusion}  \bvol{20},  \pg{655--664}.

\bibitem[{Drake} {\em et~al.\/}(1981){Drake}, {Casper}, {Clauser}, {Coensgen},
  {Correll}, {Cummins}, {Davis}, {Foote}, {Futch} \& {Goodman}]{drake1981}
{\sc \au{{Drake}, R.~P.}, \au{{Casper}, T.~A.}, \au{{Clauser}, J.~F.},
  \au{{Coensgen}, F.~H.}, \au{{Correll}, D.~L.}, \au{{Cummins}, W.~F.},
  \au{{Davis}, J.~C.}, \au{{Foote}, J.~H.}, \au{{Futch}, A.~H.} \&
  \au{{Goodman}, R.~K.}} \yr{1981}  \at{{The effect of end-cell stability on
  the confinement of the central-cell plasma in TMX}}.  \jt{Nuclear Fusion}
  \bvol{27},  \pg{359--364}.

\bibitem[Elwasif {\em et~al.\/}(2010)Elwasif, Bernholdt, Shet, Foley, Bramley,
  Batchelor \& Berry]{elwasif2010}
{\sc \au{Elwasif, W.~R.}, \au{Bernholdt, D.~E.}, \au{Shet, A.~G.}, \au{Foley,
  S.~S.}, \au{Bramley, R.}, \au{Batchelor, D.~B.} \& \au{Berry, L.~A.}}
  \yr{2010} The design and implementation of the swim integrated plasma
  simulator.  \bt{In {\em Proceedings of the 18th Euromicro Conference on
  Parallel, Distributed and Network-based Processing\/} (ed. \ed{Marco
  {Danelutto}, Julien {Bourgeois} \& Tom {Gross}})},  \pg{pp. 419--427}. IEEE,
  \publ{Computer Society Press}.

\bibitem[{Endrizzi} {\em et~al.\/}(2023){Endrizzi}, {Anderson}, {Brown},
  {Egedal}, {Geiger}, {Harvey}, {Ialovega}, {Kirch}, {Peterson}, {Petrov},
  {Pizzo}, {Qian}, {Sanwalka}, {Schmitz}, {Wallace}, {Yakovlev}, {Yu} \&
  {Forest}]{endrizzi2023}
{\sc \au{{Endrizzi}, D.}, \au{{Anderson}, J.~K.}, \au{{Brown}, M.},
  \au{{Egedal}, J.}, \au{{Geiger}, B.}, \au{{Harvey}, R.~W.}, \au{{Ialovega},
  M.}, \au{{Kirch}, J.}, \au{{Peterson}, E.}, \au{{Petrov}, Y.~V.},
  \au{{Pizzo}, J.}, \au{{Qian}, T.}, \au{{Sanwalka}, K.}, \au{{Schmitz}, O.},
  \au{{Wallace}, J.}, \au{{Yakovlev}, D.}, \au{{Yu}, M.} \& \au{{Forest},
  C.~B.}} \yr{2023}  \at{{Physics basis for the Wisconsin HTS Axisymmetric
  Mirror (WHAM)}}.  \jt{Journal of Plasma Physics}  \bvol{89}~(5),
  \pg{975890501}.

\bibitem[{Ferraro} {\em et~al.\/}(1987){Ferraro}, {Littlejohn}, {Sanuki} \&
  {Fried}]{ferraro1987}
{\sc \au{{Ferraro}, R.~D.}, \au{{Littlejohn}, R.~G.}, \au{{Sanuki}, H.} \&
  \au{{Fried}, B.~D.}} \yr{1987}  \at{{Nonlocal effects on the drift cyclotron
  loss cone dispersion relation in cylindrical geometry}}.  \jt{Physics of
  Fluids}  \bvol{30}~(4),  \pg{1115--1122}.

\bibitem[{Ferron} \& {Wong}(1984)]{ferron1984}
{\sc \au{{Ferron}, J.~R.} \& \au{{Wong}, A.~Y.}} \yr{1984}  \at{{The dependence
  of the drift cyclotron loss cone instability on the radial density
  gradient}}.  \jt{Physics of Fluids}  \bvol{27}~(5),  \pg{1287--1300}.

\bibitem[{Forest} {\em et~al.\/}(2024){Forest}, {Anderson}, {Endrizzi},
  {Egedal}, {Frank}, {Furlong}, {Ialovega}, {Kirch}, {Harvey}, {Lindley},
  {Petrov}, {Pizzo}, {Qian}, {Sanwalka}, {Schmitz}, {Wallace}, {Yakovlev} \&
  {Yu}]{forest2024}
{\sc \au{{Forest}, C.~B.}, \au{{Anderson}, J.~K.}, \au{{Endrizzi}, D.},
  \au{{Egedal}, J.}, \au{{Frank}, S.}, \au{{Furlong}, K.}, \au{{Ialovega}, M.},
  \au{{Kirch}, J.}, \au{{Harvey}, R.~W.}, \au{{Lindley}, B.}, \au{{Petrov},
  Y.~V.}, \au{{Pizzo}, J.}, \au{{Qian}, T.}, \au{{Sanwalka}, K.},
  \au{{Schmitz}, O.}, \au{{Wallace}, J.}, \au{{Yakovlev}, D.} \& \au{{Yu}, M.}}
  \yr{2024}  \at{{Prospects for a high-field, compact Break Even Axisymmetric
  Mirror (BEAM) and applications}}.  \jt{Journal of Plasma Physics}
  \bvol{90}~(1),  \pg{975900101}.

\bibitem[{Fowler} {\em et~al.\/}(2017){Fowler}, {Moir} \&
  {Simonen}]{fowler2017}
{\sc \au{{Fowler}, T.~K.}, \au{{Moir}, R.~W.} \& \au{{Simonen}, T.~C.}}
  \yr{2017}  \at{{A new simpler way to obtain high fusion power gain in tandem
  mirrors}}.  \jt{Nuclear Fusion}  \bvol{57}~(5),  \pg{056014}.

\bibitem[{Frank} {\em et~al.\/}(2024){Frank}, {Viola}, {Petrov}, {Anderson},
  {Bindl}, {Biswas}, {Caneses}, {Endrizzi}, {Furlong}, {Harvey}, {Jacobson},
  {Lindley}, {Marriott}, {Schmitz}, {Shih} \& {Forest}]{frank2024}
{\sc \au{{Frank}, S.~J.}, \au{{Viola}, J.}, \au{{Petrov}, Y.~V.},
  \au{{Anderson}, J.~K.}, \au{{Bindl}, D.}, \au{{Biswas}, B.}, \au{{Caneses},
  J.}, \au{{Endrizzi}, D.}, \au{{Furlong}, K.}, \au{{Harvey}, R.~W.},
  \au{{Jacobson}, C.~M.}, \au{{Lindley}, B.}, \au{{Marriott}, E.},
  \au{{Schmitz}, O.}, \au{{Shih}, K.} \& \au{{Forest}, C.~B.}} \yr{2024}
  \at{{Integrated modelling of equilibrium and transport in axisymmetric
  magnetic mirror fusion devices}}.  \jt{arXiv e-prints}  \pg{p.
  arXiv:2411.06644},  \arxiv{arXiv: 2411.06644}.

\bibitem[{Gary} {\em et~al.\/}(1994){Gary}, {McKean}, {Winske}, {Anderson},
  {Denton} \& {Fuselier}]{gary1994-ic-sim}
{\sc \au{{Gary}, S.~P.}, \au{{McKean}, M.~E.}, \au{{Winske}, D.},
  \au{{Anderson}, B.~J.}, \au{{Denton}, R.~E.} \& \au{{Fuselier}, S.~A.}}
  \yr{1994}  \at{{The proton cyclotron instability and the
  anisotropy/{\ensuremath{\beta}} inverse correlation}}.  \jt{\jgr}
  \bvol{99}~(A4),  \pg{5903--5914}.

\bibitem[{Gerver}(1976)]{gerver1976}
{\sc \au{{Gerver}, M.~J.}} \yr{1976}  \at{{Stabilization of drift cyclotron
  loss cone instability with additions of small amounts of cool plasma}}.
  \jt{Physics of Fluids}  \bvol{19}~(10),  \pg{1581--1590}.

\bibitem[{Gurnett} \& {Bhattacharjee}(2017)]{gurnett2017}
{\sc \au{{Gurnett}, D.~A.} \& \au{{Bhattacharjee}, A.}} \yr{2017} {\em
  {Introduction to Plasma Physics}\/}, {Second} edn.  \publ{{Cambridge
  University Press}}.

\bibitem[{Harris}(1959)]{harris1959}
{\sc \au{{Harris}, E.~G.}} \yr{1959}  \at{{Unstable Plasma Oscillations in a
  Magnetic Field}}.  \jt{\prl}  \bvol{2}~(2),  \pg{34--36}.

\bibitem[{Hasegawa}(1978)]{hasegawa1978}
{\sc \au{{Hasegawa}, A.}} \yr{1978}  \at{{Stabilization of Drift-Cyclotron
  Loss-Cone Mode by Low-Frequency Density Fluctuations}}.  \jt{\prl}
  \bvol{40}~(14),  \pg{938--941}.

\bibitem[{Hellinger} {\em et~al.\/}(2006){Hellinger},
  {Tr{\'a}vn{\'\i}{\v{c}}ek}, {Kasper} \& {Lazarus}]{hellinger2006}
{\sc \au{{Hellinger}, P.}, \au{{Tr{\'a}vn{\'\i}{\v{c}}ek}, P.}, \au{{Kasper},
  J.~C.} \& \au{{Lazarus}, A.~J.}} \yr{2006}  \at{{Solar wind proton
  temperature anisotropy: Linear theory and WIND/SWE observations}}.  \jt{\grl}
   \bvol{33}~(9),  \pg{L09101}.

\bibitem[{Ichimura} {\em et~al.\/}(1993){Ichimura}, {Inutake}, {Katsumata},
  {Hino}, {Hojo}, {Ishii}, {Tamano} \& {Miyoshi}]{ichimura1993}
{\sc \au{{Ichimura}, M.}, \au{{Inutake}, M.}, \au{{Katsumata}, R.}, \au{{Hino},
  N.}, \au{{Hojo}, H.}, \au{{Ishii}, K.}, \au{{Tamano}, T.} \& \au{{Miyoshi},
  S.}} \yr{1993}  \at{{Relaxation of pressure anisotropy due to
  Alfv{\'e}n-ion-cyclotron fluctuations observed in
  ion-cyclotron-range-of-frequency-heated mirror plasmas}}.  \jt{\prl}
  \bvol{70}~(18),  \pg{2734--2737}.

\bibitem[{Ioffe} {\em et~al.\/}(1975){Ioffe}, {Kanaev}, {Pastukhov} \&
  {Yushmanov}]{ioffe1975}
{\sc \au{{Ioffe}, M.~S.}, \au{{Kanaev}, B.~I.}, \au{{Pastukhov}, V.~P.} \&
  \au{{Yushmanov}, E.~E.}} \yr{1975}  \at{{Stabilization of cone instability of
  collisional plasma in a mirror trap}}.  \jt{Soviet Journal of Experimental
  and Theoretical Physics}  \bvol{40}~(6),  \pg{1064--1069}.

\bibitem[{Isenberg} {\em et~al.\/}(2013){Isenberg}, {Maruca} \&
  {Kasper}]{isenberg2013}
{\sc \au{{Isenberg}, P.~A.}, \au{{Maruca}, B.~A.} \& \au{{Kasper}, J.~C.}}
  \yr{2013}  \at{{Self-consistent Ion Cyclotron Anisotropy-Beta Relation for
  Solar Wind Protons}}.  \jt{\apj}  \bvol{773}~(2),  \pg{164},  \arxiv{arXiv:
  1307.1059}.

\bibitem[{Ivanov} \& {Prikhodko}(2017)]{ivanov2017}
{\sc \au{{Ivanov}, A.~A.} \& \au{{Prikhodko}, V.~V.}} \yr{2017}  \at{{Gas
  dynamic trap: experimental results and future prospects}}.  \jt{Physics
  Uspekhi}  \bvol{60}~(5),  \pg{509}.

\bibitem[{Kadomtsev} {\em et~al.\/}(1964){Kadomtsev}, {Mikhailovskii} \&
  {Timofeev}]{kadomtsev1964}
{\sc \au{{Kadomtsev}, B.~B.}, \au{{Mikhailovskii}, A.~B.} \& \au{{Timofeev},
  A.~V.}} \yr{1964}  \at{{Negative Energy Waves in Dispersive Media}}.
  \jt{Sov. Phys. JETP}  \bvol{20}~(6),  \pg{1517}.

\bibitem[{Kanaev}(1979)]{kanaev1979}
{\sc \au{{Kanaev}, B.~I.}} \yr{1979}  \at{{Stabilization of drift loss-cone
  instability (DCI) by addition of cold ions}}.  \jt{Nuclear Fusion}
  \bvol{19},  \pg{347--359}.

\bibitem[{Katsumata} {\em et~al.\/}(1996){Katsumata}, {Ichimura}, {Inutake},
  {Hojo}, {Mase} \& {Tamano}]{katsumata1996}
{\sc \au{{Katsumata}, R.}, \au{{Ichimura}, M.}, \au{{Inutake}, M.}, \au{{Hojo},
  H.}, \au{{Mase}, A.} \& \au{{Tamano}, T.}} \yr{1996}  \at{{Eigenmode
  excitation of Alfv{\'e}n ion cyclotron instability}}.  \jt{Physics of
  Plasmas}  \bvol{3}~(12),  \pg{4489--4495}.

\bibitem[{Kennel} \& {Petschek}(1966)]{kennel1966-petschek}
{\sc \au{{Kennel}, C.~F.} \& \au{{Petschek}, H.~E.}} \yr{1966}  \at{{Limit on
  Stably Trapped Particle Fluxes}}.  \jt{\jgr}  \bvol{71},  \pg{1}.

\bibitem[{Kesner}(1973)]{kesner1973}
{\sc \au{{Kesner}, J.}} \yr{1973}  \at{{Inverse ambipolar potential in a
  magnetic mirror configuration}}.  \jt{Plasma Physics}  \bvol{15}~(6),
  \pg{577--584}.

\bibitem[{Kesner}(1980)]{kesner1980}
{\sc \au{{Kesner}, J.}} \yr{1980}  \at{{Axisymmetric sloshing-ion tandem-mirror
  plugs}}.  \jt{Nuclear Fusion}  \bvol{20},  \pg{557--562}.

\bibitem[{Koepke} {\em et~al.\/}(1986{\natexlab{{\em a\/}}}){Koepke}, {Ellis},
  {Majeski} \& {McCarrick}]{koepke1986-BRLD}
{\sc \au{{Koepke}, M.}, \au{{Ellis}, R.~F.}, \au{{Majeski}, R.~P.} \&
  \au{{McCarrick}, M.~J.}} \yr{1986{\natexlab{{\em a\/}}}}  \at{{Experimental
  observation of bounce-resonance Landau damping in an axisymmetric mirror
  plasma}}.  \jt{\prl}  \bvol{56}~(12),  \pg{1256--1259}.

\bibitem[{Koepke} {\em et~al.\/}(1986{\natexlab{{\em b\/}}}){Koepke},
  {McCarrick}, {Majeski} \& {Ellis}]{koepke1986-3D}
{\sc \au{{Koepke}, M.}, \au{{McCarrick}, M.~J.}, \au{{Majeski}, R.~P.} \&
  \au{{Ellis}, R.~F.}} \yr{1986{\natexlab{{\em b\/}}}}  \at{{Three-dimensional
  mode structure of the drift cyclotron loss-cone instability in a mirror
  trap}}.  \jt{Physics of Fluids}  \bvol{29}~(10),  \pg{3439--3444}.

\bibitem[{Koepke}(1992)]{koepke1992}
{\sc \au{{Koepke}, M.~E.}} \yr{1992}  \at{{Effects of bounce-resonance damping
  on the harmonics of a plasma microinstability}}.  \jt{Physics of Fluids B}
  \bvol{4}~(5),  \pg{1193--1198}.

\bibitem[{Kotelnikov}(2024)]{kotelnikov2024}
{\sc \au{{Kotelnikov}, I.}} \yr{2024}  \at{{On the stability of the $m=1$ rigid
  ballooning mode in a mirror trap with high-beta sloshing ions}}.  \jt{arXiv
  e-prints}  \pg{p. arXiv:2406.10488},  \arxiv{arXiv: 2406.10488}.

\bibitem[{Kotelnikov} \& {Chernoshtanov}(2018)]{kotelnikov2018}
{\sc \au{{Kotelnikov}, I.~A.} \& \au{{Chernoshtanov}, I.~S.}} \yr{2018}
  \at{{Isotopic effect in microstability of electrostatic oscillations in
  magnetic mirror traps}}.  \jt{Physics of Plasmas}  \bvol{25}~(8),
  \pg{082501}.

\bibitem[{Kotelnikov} {\em et~al.\/}(2017){Kotelnikov}, {Chernoshtanov} \&
  {Prikhodko}]{kotelnikov2017}
{\sc \au{{Kotelnikov}, I.~A.}, \au{{Chernoshtanov}, I.~S.} \& \au{{Prikhodko},
  V.~V.}} \yr{2017}  \at{{Electrostatic instabilities in a mirror trap
  revisited}}.  \jt{Physics of Plasmas}  \bvol{24}~(12),  \pg{122512}.

\bibitem[{Le} {\em et~al.\/}(2023){Le}, {Stanier}, {Yin}, {Wetherton}, {Keenan}
  \& {Albright}]{le2023-hyb}
{\sc \au{{Le}, A.}, \au{{Stanier}, A.}, \au{{Yin}, L.}, \au{{Wetherton}, B.},
  \au{{Keenan}, B.} \& \au{{Albright}, B.}} \yr{2023}  \at{{Hybrid-VPIC: An
  open-source kinetic/fluid hybrid particle-in-cell code}}.  \jt{Physics of
  Plasmas}  \bvol{30}~(6),  \pg{063902},  \arxiv{arXiv: 2305.05600}.

\bibitem[{Lindgren} {\em et~al.\/}(1976){Lindgren}, {Birdsall} \&
  {Langdon}]{lindgren1976}
{\sc \au{{Lindgren}, N.~E.}, \au{{Birdsall}, C.~K.} \& \au{{Langdon}, A.~B.}}
  \yr{1976}  \at{{Electrostatic waves in an inhomogeneous collisionless
  plasma}}.  \jt{Physics of Fluids}  \bvol{19}~(7),  \pg{1026--1034}.

\bibitem[{McCarrick} {\em et~al.\/}(1987){McCarrick}, {Booske} \&
  {Ellis}]{mccarrick1987}
{\sc \au{{McCarrick}, M.~J.}, \au{{Booske}, J.~H.} \& \au{{Ellis}, R.~F.}}
  \yr{1987}  \at{{Observations of the dependence of unstable drift cyclotron
  loss cone mode characteristics on plasma density}}.  \jt{Physics of Fluids}
  \bvol{30}~(2),  \pg{614--617}.

\bibitem[McCartney {\em et~al.\/}(2014)McCartney, Hacker \&
  Yang]{mccartney2014}
{\sc \au{McCartney, G.}, \au{Hacker, T.} \& \au{Yang, B.}} \yr{2014}
  \at{{Empowering Faculty: A Campus Cyberinfrastructure Strategy for Research
  Communities}}.  \jt{Educause Review} .

\bibitem[{Mikhailovskii} \& {Timofeev}(1963)]{mikhailovskii1963}
{\sc \au{{Mikhailovskii}, A.~B.} \& \au{{Timofeev}, A.~V.}} \yr{1963}
  \at{{Theory of Cyclotron Instability in a Non-Uniform Plasma}}.  \jt{Sov.
  Phys. JETP}  \bvol{17}~(3),  \pg{626}.

\bibitem[{Myer} \& {Simon}(1980)]{myer1980}
{\sc \au{{Myer}, R.~C.} \& \au{{Simon}, A.}} \yr{1980}  \at{{Nonlinear
  saturation of the drift cyclotron loss-cone instability}}.  \jt{Physics of
  Fluids}  \bvol{23}~(5),  \pg{963--973}.

\bibitem[{Nicks} {\em et~al.\/}(2023){Nicks}, {Putvinski} \&
  {Tajima}]{nicks2023}
{\sc \au{{Nicks}, B.~S.}, \au{{Putvinski}, S.} \& \au{{Tajima}, T.}} \yr{2023}
  \at{{Stabilization of the Alfv{\'e}n-ion cyclotron instability through short
  plasmas: Fully kinetic simulations in a high-beta regime}}.  \jt{Physics of
  Plasmas}  \bvol{30}~(10),  \pg{102108}.

\bibitem[{Peterson}(2019)]{peterson2019-phd}
{\sc \au{{Peterson}, E.~E.}} \yr{2019}  \at{{A Laboratory Model for Magnetized
  Stellar Winds}}. PhD thesis, University of Wisconsin, Madison.

\bibitem[{Petrov} \& {Harvey}(2016)]{petrov2016}
{\sc \au{{Petrov}, Y.~V.} \& \au{{Harvey}, R.~W.}} \yr{2016}  \at{{A
  fully-neoclassical finite-orbit-width version of the CQL3D Fokker-Planck
  code}}.  \jt{Plasma Physics and Controlled Fusion}  \bvol{58}~(11),
  \pg{115001}.

\bibitem[{Piterski{\v{i}}} {\em et~al.\/}(1995){Piterski{\v{i}}}, {Yushmanov}
  \& {Yakovets}]{piterskii1995}
{\sc \au{{Piterski{\v{i}}}, V.~V.}, \au{{Yushmanov}, E.~E.} \& \au{{Yakovets},
  A.~N.}} \yr{1995}  \at{{Stabilization of the drift-cone instability by a flow
  shear}}.  \jt{Soviet Journal of Experimental and Theoretical Physics Letters}
   \bvol{62},  \pg{303}.

\bibitem[{Post}(1987)]{post1987-review}
{\sc \au{{Post}, R.~F.}} \yr{1987}  \at{{The magnetic mirror approach to
  fusion}}.  \jt{Nuclear Fusion}  \bvol{27}~(10),  \pg{1579}.

\bibitem[{Post} \& {Rosenbluth}(1966)]{post1966}
{\sc \au{{Post}, R.~F.} \& \au{{Rosenbluth}, M.~N.}} \yr{1966}
  \at{{Electrostatic Instabilities in Finite Mirror-Confined Plasmas}}.
  \jt{Physics of Fluids}  \bvol{9}~(4),  \pg{730--749}.

\bibitem[{Prikhodko} {\em et~al.\/}(2018){Prikhodko}, {Bagryansky},
  {Gospodchikov}, {Lizunov}, {Konshin}, {Korobeynikova}, {Kovalenko},
  {Maximov}, {Murakhtin}, {Pinzhenin}, {Savkin}, {Shalashov}, {Soldatkina},
  {Solomakhin} \& {Yakovlev}]{prikhodko2018}
{\sc \au{{Prikhodko}, V.~V.}, \au{{Bagryansky}, P.~A.}, \au{{Gospodchikov},
  E.~D.}, \au{{Lizunov}, A.~A.}, \au{{Konshin}, Z.~E.}, \au{{Korobeynikova},
  O.~A.}, \au{{Kovalenko}, Y.~V.}, \au{{Maximov}, V.~V.}, \au{{Murakhtin},
  S.~V.}, \au{{Pinzhenin}, E.~I.}, \au{{Savkin}, V.~Y.}, \au{{Shalashov},
  A.~G.}, \au{{Soldatkina}, E.~I.}, \au{{Solomakhin}, A.~L.} \& \au{{Yakovlev},
  D.~V.}} \yr{2018} {Stability and Confinement Studies in the Gas Dynamic
  Trap}.  \bt{In {\em 27th IAEA Fusion Energy Conference, Gandhinagar,
  India\/}},  \pg{pp. IAEA--CN--258/EX/P5--25}.

\bibitem[{Qian} {\em et~al.\/}(2023){Qian}, {Anderson}, {Endrizzi}, {Forest},
  {Pizzo}, {Sanwalka}, {Yakovlev}, {Yu} \& {Zarnstorff}]{qian2023}
{\sc \au{{Qian}, T.}, \au{{Anderson}, J.~K.}, \au{{Endrizzi}, D.},
  \au{{Forest}, C.~B.}, \au{{Pizzo}, J.~D.}, \au{{Sanwalka}, K.},
  \au{{Yakovlev}, D.}, \au{{Yu}, M.} \& \au{{Zarnstorff}, M.~C.}} \yr{2023}
  {Experimental Plans for MHD Stability in WHAM}.  \bt{In {\em APS Division of
  Plasma Physics Meeting Abstracts\/}},  \st{APS Meeting Abstracts},  \vol{vol.
  2023},  \pg{p. UP11.00077}.

\bibitem[{Roberts} \& {Taylor}(1962)]{roberts1962}
{\sc \au{{Roberts}, K.~V.} \& \au{{Taylor}, J.~B.}} \yr{1962}
  \at{{Magnetohydrodynamic Equations for Finite Larmor Radius}}.  \jt{\prl}
  \bvol{8}~(5),  \pg{197--198}.

\bibitem[{Rose} {\em et~al.\/}(2006){Rose}, {Genoni}, {Welch}, {Mehlhorn},
  {Porter} \& {Ditmire}]{rose2006}
{\sc \au{{Rose}, D.~V.}, \au{{Genoni}, T.~C.}, \au{{Welch}, D.~R.},
  \au{{Mehlhorn}, T.~A.}, \au{{Porter}, J.~L.} \& \au{{Ditmire}, T.}} \yr{2006}
   \at{{Flute instability growth on a magnetized plasma column}}.  \jt{Physics
  of Plasmas}  \bvol{13}~(9),  \pg{092507}.

\bibitem[{Rosenbluth}(1982)]{rosenbluth1982}
{\sc \au{{Rosenbluth}, M.~N.}} \yr{1982}  \at{{Topics in Plasma Instabilities:
  Trapped-Particle Modes and MHD}}.  \jt{Physica Scripta Volume T}  \bvol{2A},
  \pg{104--109}.

\bibitem[{Rosenbluth} {\em et~al.\/}(1962){Rosenbluth}, {Krall} \&
  {Rostoker}]{rosenbluth1962}
{\sc \au{{Rosenbluth}, M.~N.}, \au{{Krall}, N.~A.} \& \au{{Rostoker}, N.}}
  \yr{1962} {Finite Larmor Radius Stabilization of "Weakly" Unstable Confined
  Plasmas}.  \bt{In {\em Nuclear Fusion: 1962 Supplement, Part 1\/}}, ,
  \vol{vol. CN-10/170}.

\bibitem[{Ryutov} {\em et~al.\/}(2011){Ryutov}, {Berk}, {Cohen}, {Molvik} \&
  {Simonen}]{ryutov2011}
{\sc \au{{Ryutov}, D.~D.}, \au{{Berk}, H.~L.}, \au{{Cohen}, B.~I.},
  \au{{Molvik}, A.~W.} \& \au{{Simonen}, T.~C.}} \yr{2011}
  \at{{Magneto-hydrodynamically stable axisymmetric mirrorsa)}}.  \jt{Physics
  of Plasmas}  \bvol{18}~(9),  \pg{092301}.

\bibitem[{Sanuki} \& {Ferraro}(1986)]{sanuki1986}
{\sc \au{{Sanuki}, H.} \& \au{{Ferraro}, R.~D.}} \yr{1986}  \at{{Nonlocal
  Theory of DCLC Modes in a Plasma Slab with an Ambipolar Field}}.
  \jt{\physscr}  \bvol{34}~(1),  \pg{58--62}.

\bibitem[{Shmigelsky} {\em et~al.\/}(2024){Shmigelsky}, {Meyster},
  {Chernoshtanov}, {Lizunov}, {Solomakhin} \& {Yakovlev}]{shmigelsky2024-dclc}
{\sc \au{{Shmigelsky}, E.~A.}, \au{{Meyster}, A.~K.}, \au{{Chernoshtanov},
  I.~S.}, \au{{Lizunov}, A.~A.}, \au{{Solomakhin}, A.~L.} \& \au{{Yakovlev},
  D.~V.}} \yr{2024}  \at{{Kinetic instabilities in two-isotopic plasma in the
  GDT magnetic mirror}}.  \jt{{Journal of Plasma Physics}}  \bvol{90}~(6),
  \pg{905900605}.

\bibitem[Simonen {\em et~al.\/}(2008)Simonen, Cohen, Correll, Fowler, Post,
  Berk, Horton, Hooper, Fisch, Hassam, Baldwin, Pearlstein, Logan, Turner,
  Moir, Molvik, Ryutov, Ivanov, Kesner, Cohen, McLean, Tamano, Tang \&
  Imai]{simonen2008}
{\sc \au{Simonen, T.}, \au{Cohen, R.}, \au{Correll, D.}, \au{Fowler, K.},
  \au{Post, D.}, \au{Berk, H.}, \au{Horton, W.}, \au{Hooper, E.~B.}, \au{Fisch,
  N.}, \au{Hassam, A.}, \au{Baldwin, D.}, \au{Pearlstein, D.}, \au{Logan, G.},
  \au{Turner, B.}, \au{Moir, R.}, \au{Molvik, A.}, \au{Ryutov, D.}, \au{Ivanov,
  A.~A.}, \au{Kesner, J.}, \au{Cohen, B.}, \au{McLean, H.}, \au{Tamano, T.},
  \au{Tang, X.~Z.} \& \au{Imai, T.}} \yr{2008}  \bt{The axisymmetric tandem
  mirror: A magnetic mirror concept game changer magnet mirror status study
  group}. {\em Tech. Rep.\/} LLNL-TR-408176.  \org{Lawrence Livermore National
  Laboratory}.

\bibitem[{Simonen} {\em et~al.\/}(1983){Simonen}, {Allen}, {Casper}, {Clauser},
  {Clower}, {Coensgen}, {Correll}, {Cummins}, {Damm}, {Flammer}, {Foote},
  {Goodman}, {Grubb}, {Hooper}, {Hornady} \& {et al.}]{simonen1983}
{\sc \au{{Simonen}, T.~C.}, \au{{Allen}, S.~L.}, \au{{Casper}, T.~A.},
  \au{{Clauser}, J.~F.}, \au{{Clower}, C.~A.}, \au{{Coensgen}, F.~H.},
  \au{{Correll}, D.~L.}, \au{{Cummins}, W.~F.}, \au{{Damm}, C.~C.},
  \au{{Flammer}, M.}, \au{{Foote}, J.~H.}, \au{{Goodman}, R.~K.}, \au{{Grubb},
  D.~P.}, \au{{Hooper}, E.~B.}, \au{{Hornady}, R.~S.} \& \au{{et al.}}}
  \yr{1983}  \at{{Operation of the tandem-mirror plasma experiment with skew
  neutral-beam injection}}.  \jt{\prl}  \bvol{50}~(21),  \pg{1668--1671}.

\bibitem[Song {\em et~al.\/}(2022)Song, Smith, Kalyanam, Zhu, Adams, Colby,
  Finnegan, Gough, Hillery, Irvine, Maji \& St.~John]{song2022}
{\sc \au{Song, X.~C.}, \au{Smith, P.}, \au{Kalyanam, R.}, \au{Zhu, X.},
  \au{Adams, E.}, \au{Colby, K.}, \au{Finnegan, P.}, \au{Gough, E.},
  \au{Hillery, E.}, \au{Irvine, R.}, \au{Maji, A.} \& \au{St.~John, J.}}
  \yr{2022} Anvil - system architecture and experiences from deployment and
  early user operations.  \bt{In {\em Practice and Experience in Advanced
  Research Computing 2022: Revolutionary: Computing, Connections, You\/}}, {\em
  PEARC '22\/} .  \publ{New York, NY, USA: Association for Computing
  Machinery}.

\bibitem[{Stanier} {\em et~al.\/}(2019){Stanier}, {Chac{\'o}n} \&
  {Chen}]{stanier2019-pic}
{\sc \au{{Stanier}, A.}, \au{{Chac{\'o}n}, L.} \& \au{{Chen}, G.}} \yr{2019}
  \at{{A fully implicit, conservative, non-linear, electromagnetic hybrid
  particle-ion/fluid-electron algorithm}}.  \jt{Journal of Computational
  Physics}  \bvol{376},  \pg{597--616},  \arxiv{arXiv: 1803.07158}.

\bibitem[{Stix}(1992)]{stix1992}
{\sc \au{{Stix}, T.~H.}} \yr{1992} {\em {Waves in Plasmas}\/}.  \publ{Melville,
  New York: American Institute of Physics}.

\bibitem[{Tajima} \& {Mima}(1980)]{tajima1980}
{\sc \au{{Tajima}, T.} \& \au{{Mima}, K.}} \yr{1980}  \at{{Stabilization of the
  Alfv{\'e}n-ion cyclotron instability in inhomogeneous media}}.  \jt{Physics
  of Fluids}  \bvol{23}~(3),  \pg{577--589}.

\bibitem[{Tajima} {\em et~al.\/}(1977){Tajima}, {Mima} \& {Dawson}]{tajima1977}
{\sc \au{{Tajima}, T.}, \au{{Mima}, K.} \& \au{{Dawson}, J.~M.}} \yr{1977}
  \at{{Alfv{\'e}n Ion-Cyclotron Instability: Its Physical Mechanism and
  Observation in Computer Simulation}}.  \jt{\prl}  \bvol{39}~(4),
  \pg{201--204}.

\bibitem[{Tang} {\em et~al.\/}(1972){Tang}, {Pearlstein} \& {Berk}]{tang1972}
{\sc \au{{Tang}, W.~M.}, \au{{Pearlstein}, L.~D.} \& \au{{Berk}, H.~L.}}
  \yr{1972}  \at{{Finite Beta Stabilization of the Drift-Cone Instability}}.
  \jt{Physics of Fluids}  \bvol{15}~(6),  \pg{1153--1155}.

\bibitem[{Tsidulko} \& {Chernoshtanov}(2014)]{tsidulko2014}
{\sc \au{{Tsidulko}, Y.~A.} \& \au{{Chernoshtanov}, I.~S.}} \yr{2014}
  \at{{Alfv{\'e}n ion-cyclotron instability in an axisymmetric trap with
  oblique injection of fast atoms}}.  \jt{Plasma Physics Reports}
  \bvol{40}~(12),  \pg{955--964}.

\bibitem[{Watson}(1922)]{watson1922}
{\sc \au{{Watson}, G.~N.}} \yr{1922} {\em {A Treatise on the Theory of Bessel
  Functions}\/}.  \publ{{Cambridge University Press}}.

\bibitem[{Wetherton} {\em et~al.\/}(2021){Wetherton}, {Le}, {Egedal}, {Forest},
  {Daughton}, {Stanier} \& {Boldyrev}]{wetherton2021}
{\sc \au{{Wetherton}, B.~A.}, \au{{Le}, A.}, \au{{Egedal}, J.}, \au{{Forest},
  C.}, \au{{Daughton}, W.}, \au{{Stanier}, A.} \& \au{{Boldyrev}, S.}}
  \yr{2021}  \at{{A drift kinetic model for the expander region of a magnetic
  mirror}}.  \jt{Physics of Plasmas}  \bvol{28}~(4),  \pg{042510},
  \arxiv{arXiv: 2105.01572}.

\bibitem[{Yakovlev} {\em et~al.\/}(2018){Yakovlev}, {Shalashov},
  {Gospodchikov}, {Maximov}, {Prikhodko}, {Savkin}, {Soldatkina}, {Solomakhin}
  \& {Bagryansky}]{yakovlev2018}
{\sc \au{{Yakovlev}, D.~V.}, \au{{Shalashov}, A.~G.}, \au{{Gospodchikov},
  E.~D.}, \au{{Maximov}, V.~V.}, \au{{Prikhodko}, V.~V.}, \au{{Savkin}, V.~Y.},
  \au{{Soldatkina}, E.~I.}, \au{{Solomakhin}, A.~L.} \& \au{{Bagryansky},
  P.~A.}} \yr{2018}  \at{{Stable confinement of high-electron-temperature
  plasmas in the GDT experiment}}.  \jt{Nuclear Fusion}  \bvol{58}~(9),
  \pg{094001}.

\bibitem[{Yamaguchi}(1996)]{yamaguchi1996}
{\sc \au{{Yamaguchi}, H.}} \yr{1996}  \at{{Amplitude Oscillation of an
  Instability in a Plasma in the Presence of a Damping Mechanism}}.
  \jt{Journal of the Physical Society of Japan}  \bvol{65}~(10),  \pg{3115}.

\bibitem[{Yerger} {\em et~al.\/}(2025){Yerger}, {Kunz}, {Bott} \&
  {Spitkovsky}]{yerger2025}
{\sc \au{{Yerger}, E.~L.}, \au{{Kunz}, M.~W.}, \au{{Bott}, A. F.~A.} \&
  \au{{Spitkovsky}, A.}} \yr{2025}  \at{{Collisionless conduction in a
  high-beta plasma: a collision operator for whistler turbulence}}.
  \jt{Journal of Plasma Physics}  \bvol{91}~(1),  \pg{E20},  \arxiv{arXiv:
  2405.06481}.

\bibitem[{Yoshikawa} {\em et~al.\/}(2019){Yoshikawa}, {Kohagura}, {Ikezoe},
  {Sakamoto}, {Ezumi}, {Minami}, {Hirata}, {Terakado}, {Nojiri}, {Jang},
  {Tanaka}, {Ichimura}, {Imai} \& {Nakashima}]{yoshikawa2019}
{\sc \au{{Yoshikawa}, M.}, \au{{Kohagura}, J.}, \au{{Ikezoe}, R.},
  \au{{Sakamoto}, M.}, \au{{Ezumi}, N.}, \au{{Minami}, R.}, \au{{Hirata}, M.},
  \au{{Terakado}, A.}, \au{{Nojiri}, K.}, \au{{Jang}, S.}, \au{{Tanaka}, A.},
  \au{{Ichimura}, M.}, \au{{Imai}, T.} \& \au{{Nakashima}, Y.}} \yr{2019}
  \at{{Suppression of flute-like fluctuation by potential formation in GAMMA
  10/PDX}}.  \jt{Nuclear Fusion}  \bvol{59}~(7),  \pg{076031}.

\end{thebibliography}

\appendix

\newpage
\pagebreak
\section{Hyper-Resistivity Scan} \label{app:hypereta}

\begin{figure*}
    \centering
    \includegraphics[width=\textwidth]{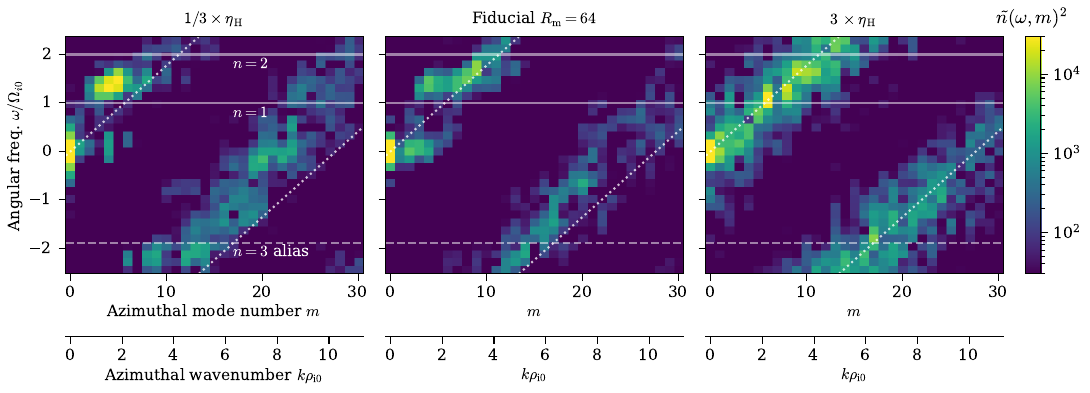}
    \caption{
        Effect of hyper-resistivity, increasing left to right, upon
        density-fluctuation Fourier spectra in WHAM $\mratio=64$ simulations;
        center panel is same data as Figure~\ref{fig:mode-f}(c).
        Fourier spectra and annotations constructed like in
        Figure~\ref{fig:mode-f}, at same $r=2.69\rLio$ over $t=3$ to
        $6\,\tbounce$, but the colormap range and 2D plot domain/range are
        changed.
    }
    \label{fig:mode-f-hypereta}
\end{figure*}

Hyper-resistivity, although intended to suppress grid-scale numerical noise,
may also alter DCLC mode structure as discussed in {\S}\ref{sec:methods-geometry}.
Figure~\ref{fig:mode-f-hypereta} shows that raising or decreasing
$\eta_\mt{H}$ by a factor of $3$ alters the spectrum of density fluctuations at
the plasma edge; with higher $\eta_\mt{H}$, the spectrum broadens and is less
coherent.

\newpage
\pagebreak
\section{Electron Parallel Response} \label{app:electrons}

To obtain the electrons' parallel response in \S\ref{sec:ELD}, we start
again from \citet{stix1992}, \S14-3, Eq.~(8), neglecting spatial
derivatives of order $\ptl^2 g/\ptl y^2$ or higher in Stix's notation,
where $g$ is the guiding-center distribution.
For a Maxwellian guiding-center distribution, the susceptibility is:
\begin{align} \label{eq:chi-harris-gradient}
    \chi_\mt{s} = \left(\frac{\omps}{\Omcs}\right)^2
    \sum_{n=-\infty}^{\infty}
    e^{-\lambda}
    I_n(\lambda)
    \bigg\{
        &\left(\frac{k_\perp}{k}\right)^2
        \left[
            \frac{2 n}{k_\perp^2}
            \left( 1 - \frac{n \epsilon}{k_\perp} \right)
            +
            \frac{\epsilon}{k_\perp}
        \right]
        \frac{Z(\zeta_n)}{k_\prll} \nonumber \\
        &+\left(\frac{k_\parallel}{k}\right)^2
        \frac{2}{k_\prll^2}
        \left( 1 - \frac{n \epsilon}{k_\perp} \right)
        \left[
            1 +  \zeta_n Z(\zeta_n)
        \right]
    \bigg\}
    \, ,
\end{align}
where the modified Bessel function
$I_n(\lambda)$ has argument $\lambda = k_\perp^2/2$,
the plasma dispersion function $Z(\zeta_n)$
has argument $\zeta_n = (\omega-n)/k_\prll$,
and the variables $\omega, \epsilon, k, k_\perp, k_\prll$ are all
dimensionless following the same \emph{species-specific} scheme used for
Equation~\eqref{eq:chi-general}.
Equation~\eqref{eq:chi-harris-gradient} is valid for any $\vec{k}$ angle.
The limit $\zeta_n \to \infty$ and $k_\prll/k \to 0$ recovers the
perpendicular susceptibility given by Equation~\eqref{eq:chi-general}.
The limit $\epsilon/k_\perp \to 0$ recovers the standard \citet{harris1959}
dispersion relation \citep[{\S}10.2]{gurnett2017}.

Let us simplify Equation~\eqref{eq:chi-harris-gradient}.
Take the limits
$\lambda \to 0$
and
$\zeta_{\pm 1} = (\omega \mp 1)/k_\prll \approx \mp 1/k_\prll \to \infty$
in order to expand
$e^{-\lambda} I_n(\lambda)$
and $Z(\zeta_{\pm 1})$.
But, make no assumptions on the magnitude of $\zeta_0 = \omega/k_\prll$.
Also, drop all $|n| > 1$ Bessel terms.
The result is:
\begin{equation} \label{eq:chi-PKPM}
    \chi_\mt{s} = \left(\frac{\omps}{\Omcs}\right)^2
    \left\{
        \left(\frac{k_\perp}{k}\right)^2
        \left[
            \frac{\epsilon}{k_\perp \omega}
            - 1
        \right]
        \zeta_0 Z(\zeta_0)
        + \left(\frac{k_\parallel}{k}\right)^2
        \frac{2}{k_\prll^2}
        \left[ 1 +  \zeta_0 Z(\zeta_0) \right]
    \right\}
    \, .
\end{equation}
which yields Equation~\eqref{eq:chi-PKPM-maintext} after putting in dimensions.
The limit $\zeta_0 \to \infty$ recovers the familiar cold-fluid result,
written below in dimension-\emph{ful} variables:
\begin{equation} \label{eq:chi-PCPM}
    \chi_\mt{s} =
        \left(\frac{k_\perp}{k}\right)^2
        \frac{\omps^2}{\Omcs^2}
        \left[
            1
            - \frac{\epsilon \Omcs}{k_\perp \omega}
        \right]
        - \left(\frac{k_\prll}{k}\right)^2
        \frac{\omps^2}{\omega^2}
    \, .
\end{equation}

A subtlety appears when expanding Equation~\eqref{eq:chi-harris-gradient} into
Equation~\eqref{eq:chi-PKPM}.
Consider the susceptibility tensor components $\chi_{\perp\perp}$ and
$\chi_{\prll\prll}$ for a hot homogeneous plasma \citep[{\S}10]{stix1992}.
The perpendicular response simplifies in the cold-fluid limit:
\begin{equation} \label{eq:app-chixx-cold}
    \chi_{\perp\perp} \to \frac{\omps^2}{\Omcs^2}
    \, .
\end{equation}
But, $\chi_{\perp\perp}$ is \emph{cancelled} by the analogous expansion of
$\chi_{\prll\prll}$ when (i) both $k_\perp$ and $k_\prll$ are
finite, (ii) $\zeta_0$ is kept finite, and (iii) terms of order
$\mathcal{O}(\lambda^1)$ are kept in expanding $e^{-\lambda} I_0(\lambda)$.
Said expansion gives:
\begin{align} \label{eq:app-chizz-cold}
    \chi_{\prll\prll}
    &\to
    \frac{\omps^2}{\Omcs^2}
    \frac{2}{k_\prll^2} \left[ 1 + \zeta_0 Z(\zeta_0) \right]
        \left( 1 - \lambda + \mathcal{O}(\lambda^2) \right)
    \nonumber \\
    &=
    \frac{\omps^2}{\Omcs^2}
    \left\{
    \frac{2}{k_\prll^2} \left[ 1 + \zeta_0 Z(\zeta_0) \right]
      - \frac{k_\perp^2}{k_\prll^2}
      - \frac{k_\perp^2}{k_\prll^2} \zeta_0 Z(\zeta_0)
    \right\}
    \, ,
\end{align}
with $k_\prll$ dimensionless as before.
Then, in the combined electrostatic susceptibility
\begin{equation} \label{eq:app-chi-sum}
    \chi
    \approx (k_\perp/k)^2 \chi_{\perp\perp}
    + (k_\prll/k)^2 \chi_{\prll\prll}
    + (2 k_\prll k_\perp/k^2) \chi_{\perp\prll}
    \, ,
\end{equation}
we see that Equations~\eqref{eq:app-chixx-cold} and \eqref{eq:app-chizz-cold}
partly cancel, and only a \emph{parallel} contribution
$-(k_\perp/k)^2 (\omps/\Omcs)^2 \zeta_0 Z(\zeta_0)$ remains.
This remainder term can be seen in Equation~\eqref{eq:chi-PKPM}.
When the $\zeta_0 \to \infty$ limit is taken, it is this parallel
remainder that provides the usual perpendicular response
$\chi_{\perp\perp} \to \omps^2/\Omcs^2$.
This is to some extent a semantic quibble; we can also say that the
remainder term $-(k_\perp/k)^2 (\omps/\Omcs)^2 \zeta_0 Z(\zeta_0)$ cancels
the parallel term $-(k_\perp/k)^2 (\omps/\Omcs)^2$ to leave only the
perpendicular term $+(k_\perp/k)^2 (\omps/\Omcs)^2$ in
Equation~\eqref{eq:app-chi-sum}.
But, the overall point stands that the perpendicular term in
Equation~\eqref{eq:chi-PKPM} can be significantly modified by the parallel
response, even when $k_\parallel \ll k_\perp$; the regulating parameter is
$\zeta_0$.

\end{document}